\documentclass{aastex631} 
\usepackage{dirtytalk}

\usepackage{booktabs} 
\usepackage{graphicx} 
\usepackage{multirow}
\usepackage{algorithmic}
\usepackage{algorithm}
\usepackage{float}
\definecolor{darkgreen}{RGB}{0,100,0} 
\usepackage{array}
\usepackage{amsmath}
\newcolumntype{C}[1]{>{\centering\arraybackslash}p{#1}}

\usepackage{hyperref}
\hypersetup{
    pdftitle={},        
    pdfauthor={},       
    pdfsubject={},      
    pdfkeywords={},     
    pdfproducer={},     
    pdfcreator={}       
}

\begin{document}

\title{Solar Active Regions Emergence Prediction Using Long Short-Term Memory Networks}





\author[0000-0002-0786-7307]{Spiridon Kasapis}
\affiliation{NASA Ames Research Center, Moffett Field, CA, USA}

\author[0000-0003-4144-2270]{Irina N. Kitiashvili}
\affiliation{NASA Ames Research Center, Moffett Field, CA, USA}

\author[0000-0003-0364-4883]{Alexander G. Kosovichev}
\affiliation{New Jersey Institute of Technology, Newark, NJ 07103, USA}
\affiliation{NASA Ames Research Center, Moffett Field, CA, USA}

\author{John T. Stefan}
\affiliation{New Jersey Institute of Technology, Newark, NJ 07103, USA}

\begin{abstract}

We developed Long Short-Term Memory (LSTM) models to predict the formation of active regions (ARs) on the solar surface. Using the Doppler shift velocity, the continuum intensity, and the magnetic field observations from the Solar Dynamics Observatory (SDO) Helioseismic and Magnetic Imager (HMI), we have created time-series datasets of acoustic power and magnetic flux, which are used to train LSTM models on predicting continuum intensity, 12 hours in advance. These novel machine learning (ML) models are able to capture variations of the acoustic power density associated with upcoming magnetic flux emergence and continuum intensity decrease. Testing of the models' performance was done on data for 5 ARs, unseen from the models during training. Model 8, the best performing model trained, was able to make a successful prediction of emergence for all testing active regions in an experimental setting and three of them in an operational. The model predicted the emergence of AR11726, AR13165, and AR13179 respectively 10, 29, and 5 hours in advance, and variations of this model achieved average RMSE values of 0.11 for both active and quiet areas on the solar disc. This work sets the foundations for ML-aided prediction of solar ARs.

\end{abstract}



\section{Introduction} \label{sec:intro}

The growing interest in manned spaceflight and the increased reliance on space-based infrastructure has made reliable space weather forecasting tools very important. Solar activity is the primary influencing factor of space weather and the geospace environment; therefore, understanding of its origin should form the foundation of these forecasting tools. Lately, significant efforts have been made towards the prediction of certain aspects of space weather, for example, flares \citep[e.g.,][]{georgoulis2013toward, claspflarepred}, coronal mass ejections \citep[e.g.,][]{maharana2024solar}, and solar energetic particles \citep[e.g.,][]{ali2024predicting, kasapis2024forecasting}, but forecasting the progenitor of these events, namely the solar active regions (ARs), has been a rather unexplored area of study. The main difference between forecasting space weather events and the emergence of active regions lies in their observational constraints: there are strong connections between the strength and configuration of the photospheric magnetic field and the likelihood of flares \citep{li2024survey} and CMEs, so prediction of these events benefits greatly from the high-resolution magnetograms produced by the Helioseismic and Magnetic Imager (HMI) onboard the Solar Dynamics Observatory (SDO). Forecasting the emergence of ARs, meanwhile, requires probing of the solar interior that is only possible indirectly via helioseismic methods \citep[e.g.,][]{ilonidis2011detection,stefan2023exploring}.

To detect emergence, it is important to capture those subsurface solar activity signatures that precede the abrupt decrease in continuum intensity and increase in magnetic flux during the formation of an AR. By using time-distance helioseismology to image sound-speed variations prior to the appearance of the active region on the solar surface, \cite{kosovichev2001time} made the very first attempt to observe an emerging active region in the solar interior. This research focused on increases in magnetic flux associated with the AR emergence, showing that the flux is fragmented and travels very quickly through the top 18 Mm layer of the convection zone, with a speed of about 1 km/s. Similarly, \cite{hartlep2011signatures} explored the feasibility of using variations in acoustic power as a precursor to detect signatures of emerging magnetic flux for AR10488. More specifically, they demonstrated that subsurface structures influence the acoustic power at the overlying photosphere, suggesting that changes in acoustic power could be indicative of active regions before they become visible on the solar surface. 

There is a deep body of past work aimed at using these helioseismic methods to attempt the detection of magnetic flux before it reaches the surface, with the majority of research focusing on the near-surface layers (z $\geq -20$ Mm) where the signal-to-noise ratio (SNR) is greater. Perhaps the most comprehensive of these works is the analysis of helioseismic signatures preceding the emergence of over 100 ARs \citep{birch2012helioseismology, leka2013helioseismology, barnes2014helioseismology}, where statistically significant differences in subsurface flows and wave speeds preceding active region formation were identified. In the series of works by \cite{birch2012helioseismology}, \cite{leka2013helioseismology} and  \cite{barnes2014helioseismology} the authors tested whether pre-appearance signatures of solar magnetic active regions were detectable using various tools of local helioseismology. Using data from the Global Oscillations Network Group \citep[GONG,][]{harvey1996global} and SOHO/MDI \citep{scherrer1995solar}, this series of studies a) ruled out spatially extended flows above 15~m/s in the top 20 Mm of the photosphere before emergence, thereby setting constraints on theoretical models of active region development and b) presented evidence of a helioseismic precursor to AR emergence that is present for at least a day prior to emergence. 

The possibility of detection of large active regions before they become visible using synoptic imaging of subsurface magnetic activity was also demonstrated by \cite{ilonidis2011detection}, who detected strong acoustic travel-time anomalies of an order of 12 to 16 seconds as deep as 65~Mm. \cite{ilonidis2013helioseismic} proved that originating from deeper layers, these acoustic anomalies rise to shallower regions at velocities up to 1~km/s, suggesting their association with acoustic power variations rather than just subsurface flows or wave-speed perturbations. The deviations in the mean phase travel time of acoustic waves before the emergence of 46 large active regions was investigated by \cite{stefan2023exploring}, showing the relationship between subsurface acoustic signals and surface magnetic flux. In a similar vein, \cite{attie2018precursors} observed disruption of the moat flow near active region AR12673, several hours before the onset of strong flux emergence. This disruption occurred exactly where magnetic flux would later emerge, identifying horizontal divergent flows at the solar surface as potential precursors to this flux. 

These divergent and convergent flows that serve as precursors to the emergence of ARs have been extensively studied \citep[e.g.,][]{rees2022preemergence,gottschling2021evolution,schunker2024flux}. Therefore, to explore the complex subsurface processes preceding AR emergence from different vantage points, more recent studies have concentrated on the strengthening of the solar f-mode prior to emergence \citep{waidele2023strengthening,singh2016high}. New calibration techniques used by \cite{korpi2022solar} did not reveal a significant enhancement of the f-mode prior to AR emergence. This absence of f-mode variations, especially on smaller scales, could be attributed to the limited magnetic sensitivity and spatial resolution of HMI. While significant research has been devoted to understanding the complex behavior of subsurface activity preceding AR emergence, and despite the availability of relevant datasets \citep{schunker2016sdo}, no study has yet directly addressed the challenge of predicting the emergence itself.

To address this gap, and given the evidence that subsurface variations precede the emergence of solar ARs, in this research, we use the mean acoustic power and the mean magnetic flux derived from the SDO/HMI data to train ML models that predict the decrease in continuum intensity, associated with the emergence of active regions, before it becomes visible on the solar surface. To train the models, a dataset of 61 emerging ARs was created as described in Section~\ref{sec:Dataset}. In Section ~\ref{sec:LSTM} we provide information about the Long Short-Term Memory (LSTM) models used in this research, such as their architecture and the training and testing methods employed. The AR prediction results are discussed in Sections~\ref{sec:Results} and \ref{sec:IAPs}, while in Section~\ref{sec:Conclusion} we offer some discussion on the ML model predictions along with some recommendations for future work. 


\section{Dataset} \label{sec:Dataset}

To train and test LSTM models to forecast the emergence of ARs, a novel ML-ready dataset was created by tracking before, during, and after their emergence 61 ARs, using data from the Helioseismic and Magnetic Imager \citep[HMI]{scherrer2012helioseismic} onboard the Solar Dynamics Observatory \citep[SDO]{pesnell2012solar}. Each selected AR appeared on the solar surface within 30 degrees longitude from the central meridian between March 1st, 2010 and June 1st, 2023, persisted for more than 4 days, and reached a total area of 200 millionths of the solar hemisphere. This longitude range was chosen to minimize significant distortion due to projection and center-to-limb effects. 

\begin{figure}[ht] 
\centering
\includegraphics[width=0.95\textwidth]{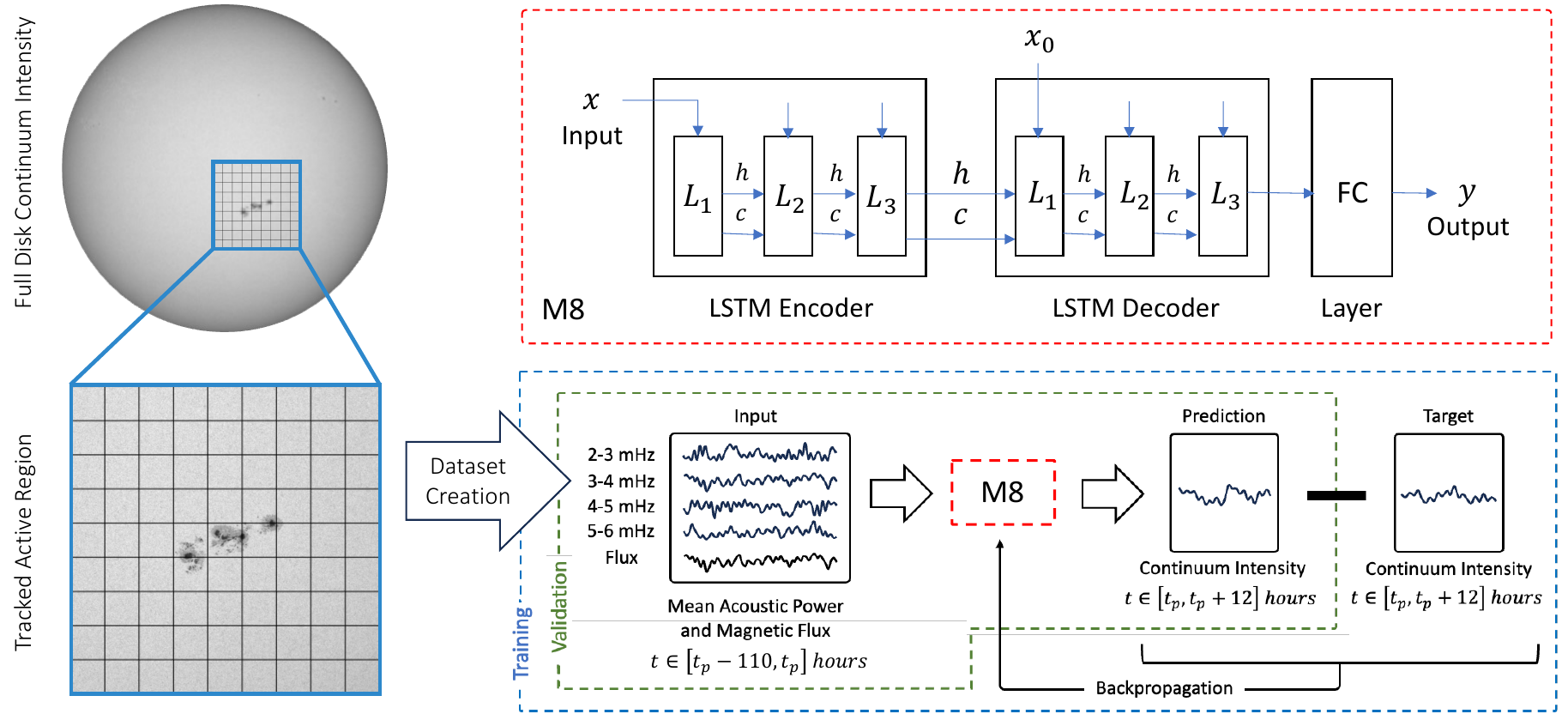}
\caption{Diagram of the research data processing sequence: 61 ARs have been tracked from the SDO/HMI full disk $V_D$, $\Phi_m$ and $I_c$ maps. The tracked regions were split into smaller tiles, and the timeline datasets were created by averaging the values of each tile. The timelines are used as inputs during the training and validation/testing process of different LSTM models, such as Model 8 (M8) seen in red. The architecture of Model 8 is presented in the area enclosed by the red dotted line.} \label{fig:Diagram}
\end{figure}

The HMI instrument provides high-resolution maps of three different physical quantities: a) the Doppler velocity $V_D$, b) the line-of-sight magnetic flux $\Phi_m$ and c) the continuum intensity $I_c$ on the solar surface. For all three data products, we tracked the corresponding $30.66\times30.66$ degrees (heliographic coordinates) area around the 61 ARs (Figure~\ref{fig:Diagram}, top left) included in the dataset, taking into account the local rotation speed. These solar disc patches represent a $512\times512$-pixel square centered around the target AR and are tracked through time by dividing them into overlapping 8-hour time series. Each one of these overlapping 8-hour series is comprised of 640 frames, with a cadence of 45 seconds. Although the $\Phi_m$ and $I_c$ maps are used without further processing, the Dopplergram ($V_D$ maps) tracked regions are used to generate acoustic power ($P_a$) maps for four frequency ranges: 2–3, 3–4, 4–5, and 5–6 mHz.

The resulting $P_a$ maps, along with the corresponding $\Phi_m$ and $I_c$ tracked regions, were split into a grid of 9 by 9 tiles (Figure \ref{fig:Diagram}, left). By splitting the tracked regions, we focus the AR evolution tracing on a local 57 by 57-pixel area (corresponding to 3.36~degrees of heliographic longitude and latitude) which is then reduced to structured timelines conducive to machine learning analysis by calculating, for each frame, the mean of each individual tile (Figure \ref{fig:Diagram}, Input Box). This ensemble of mean $P_a$, $\Phi_m$ and $I_c$ time-series is further processed by removing the solar sphere geometric effect and by normalizing them. Further details on the process used to create time-series data can be found in the Data Preparation Section of \cite{kasapis2024predicting}.

\begin{table}[h] 
\begin{tabular}{cccccc}
\toprule
  & Mean $\phi$  & Mean Traveled $\lambda$ & Mean $A_{max}$ & NOAA Record Span & Number \\ \midrule
Training ARs  & $ 9.3 \pm 16.58$ & $87.6 \pm 18.16$ & $318.0 \pm 169.91$ & $6.6 \pm 1.52$  & 40 \\ 
Testing ARs  & $-5.9 \pm 17.53$ & $79.8 \pm 15.35$ & $424.0 \pm 332.08$ & $5.8 \pm 1.10$ & 5 \\ 
Entire Dataset  & $7.6 \pm 17.17$ & $86.7 \pm 17.89$ & $329.8 \pm 191.70$ & $6.5 \pm 1.49$ & 45 \\ 
\midrule
Units & degrees  & degrees & $\dfrac{1}{10^6}$ of Hemisphere &  days &  \\ 
\bottomrule
\end{tabular}
\caption{Summary of the mean and standard deviation values of the 40 training and 5 testing ARs, for their observed latitude ($\phi$), maximum area ($A_{max}$), longitude ($\lambda$) traveled, and the NOAA record span that corresponds to this longitudinal movement.} \label{tab:DatasetComparissons}
\end{table}

Out of the 61 tracked solar ARs, 46 emerged during Solar Cycle 24 and 15 during Solar Cycle 25. Data gaps and quality issues are present in 16 of the tracked ARs, which prohibit us from using them for ML training and testing without further investigation and processing (such as interpolation). Therefore, for the research presented in the current paper, only a total of 45 ARs are used, 40 for training and 5 for testing -- almost a 90-10 train-test data ratio. Table~\ref{tab:DatasetComparissons} outlines the mean values of the ARs latitude (Mean $\phi$), traveled longitude (Mean Traveled $\lambda$), maximum area (Mean $A_{max}$), and the time of the AR's life (NOAA Record Span) on the observable disc for the training and testing datasets but also for the entire dataset (training and testing combined). The ARs in our dataset were observed on the solar disk for a period of 6.5~days on average, which corresponds to 86.7~degrees in heliographic coordinates, while the mean maximum area was 329.8~millionths of a hemisphere (mh). As mentioned, in this research, only ARs that were bigger than 200~mh were analyzed; therefore, the average AR in our dataset is small in size compared to larger ARs (such as AR11726) which can reach areas of up to 1000~mh. The mean latitude of the dataset is 7.6~heliographic degrees with a standard deviation of 17.17~degrees, showing that there is a diversity of ARs in terms of latitude, within the range of latitudes that ARs usually emerge. 

Table~\ref{tab:Testing} presents the 5 ARs that the testing dataset includes and their characteristics. The first two emerged during Solar Cycle 24 (2013), and the other three during Solar Cycle 25 (2022-2023). The selection of ARs for the training and testing datasets was balanced to ensure equal representation of the two Solar Cycles in the testing group while also maintaining similar characteristics between the two groups. Another reason why the three most recent ARs in our dataset were chosen for training is that we would like to simulate, as much as possible, an operational setting where we are trying to predict future events by training an ML model with data from the past. Similarly, Table~\ref{tab:Training} in the Appendix presents information for the 40 ARs that are used for training. Both Tables~\ref{tab:Testing} and \ref{tab:Training} include information about the time of emergence and disappearance behind the East limb, the latitude and longitude, the size of the ARs, and their classification at the time of maximum area. The coordinate units are degrees, and the size is measured in millionths of a hemisphere (Table~\ref{tab:DatasetComparissons}). For classification of the ARs in our dataset, we use the three-component McIntosh classification \citep{mcintosh1990classification} and the Mount Wilson (or Hale) classification \citep{chapman1939magnetic, jaeggli2016magnetic}. The McIntosh classification follows the format 'Zpc', where 'Z' denotes the modified Zurich Class, 'p' characterizes the penumbra of the principal spot, and 'c' describes the distribution of spots within the group's interior.

\begin{table}[ht] 
\label{tab:Testing}
\centering
\begin{tabular}{@{}ccccccccccccc@{}}
\toprule
AR\# & First Record & Last Record & $\phi$ & $\lambda_{Carr}$ & $\lambda_{s}$ & $\lambda_{e}$ & $A_s$ & $A_e$ & $A_{max}$ & $A_{max}$ Date & McIntosh & Hale  \\
\midrule
11698 & 2013.03.15 & 2013.03.19 & -19.5 & 117.0 & 29.0 & 86.0 & 20 & 200 & 200 & 2013.03.18 & Eao & $\beta$ \\
11726 & 2013.04.20 & 2013.04.27 & 13.0 & 327.0 & -7.0 & 93.0 & 20 & 600 & 1000 & 2013.04.26 & Fkc & $\beta \gamma \delta$ \\
13165 & 2022.12.12 & 2022.12.18 & -20.0 & 277.5 & 10.0 & 88.0 & 20 & 150 & 340 & 2022.12.16 & Ekc & $\beta \delta$ \\
13179 & 2022.12.30 & 2023.01.05 & 13.5 & 43.0 & 11.0 & 92.0 & 30 & 80 & 380 & 2023.01.03 & Dko & $\beta$ \\
13183 & 2023.01.06 & 2023.01.12 & -16.5 & 309.0 & 8.0 & 91.0 & 30 & 50 & 200 & 2023.01.08 & Dso & $\beta \delta$ \\
\bottomrule
\end{tabular}
\caption{List of the 5 ARs used for testing, along with information about their assigned NOAA active region number (AR\#), the date of NOAA's first (First Record) and last (Last Record) record, the constant latitude in which the AR emerged ($\phi$), the Carrington longitude ($\lambda_{Carr}$), the starting ($\lambda_{s}$) and ending ($\lambda_{e}$) longitude, the starting ($A_s$), ending ($A_e$) and maximum ($A_{max}$) area of the AR and the date of the maximum area ($A_{max}$ Date).}
\end{table} 

ARs travel at an almost constant latitude $\phi$. The latitude standard deviation (16.58 and 17.53 degrees) values are pretty high compared to the mean values (9.3 and -5.9 degrees) for the training and testing datasets; therefore, both groups have samples from a variety of different latitudes. More specifically, as seen in Table~\ref{tab:Testing}, the latitudes of the testing ARs vary from -20 to 13.5, while the mean traveled longitude in both datasets is very similar (87.6 and 79.8 degrees), with the testing dataset presenting slightly lower values, a difference reflected in the NOAA Record span too (6.6 versus 5.8 days). A shorter traveled longitude range and, therefore, shorter observed AR time on the disc for the testing dataset means that the testing ARs emerged further away from the East limb. This is important as during testing, we want an adequate amount of data (at least 110~hours, as will be discussed in Section~\ref{sec:LSTM}) before the emergence of an AR in order to perform reliable prediction. Both datasets have similar size distributions (maximum AR area $A_{max}$), with the testing ARs presenting a higher mean only because of the inclusion in the testing dataset of AR11726, the biggest AR tracked in this research. AR11726 was specifically reserved for testing because not only did it grow rapidly by reaching 1000~mh within 7~days, but it is also one AR that caused a lot of eruptive activity. Predicting impactful space weather ARs is crucial for this research. All differences between training and testing data in Table~\ref{tab:DatasetComparissons} are within the prescribed standard deviations; therefore, the two groups, despite some inevitable differences, are quite similar, ensuring the reliable testing of the trained LSTM models discussed in the next Section.


\section{LSTM Models Training, Testing and Architecture} \label{sec:LSTM}

Understanding and capturing those changes in $P_a$ and $\Phi_m$ that precede the emergence of an AR and thereby enable their prediction is a complex and challenging task. Long Short-Term Memory \citep[LSTM;][]{sherstinsky2020fundamentals} networks have been for many years a popular machine learning method used successfully for various tasks, such as flood forecasting \citep[e.g.,][]{le2019application}, sea surface temperature \citep[e.g.,][]{zhang2017prediction} and even financial market \citep[e.g.,][]{fischer2018deep} predictions. In this section, we discuss three key aspects of the LSTM models used for forecasting AR emergence: the training and testing methodologies employed and the architecture of the various networks.

We train the LSTM models using the $P_a$, $\Phi_m$, and $I_c$ time series that describe the evolution of the ARs in Tables~\ref{tab:Training} and~\ref{tab:Testing}. Algorithm \ref{alg:lstm_training} outlines the training process we follow. The first step is to load the ML-ready data, the data that have gone through the processing discussed in Section \ref{sec:LSTM} and are in the form of 240-hours-long time-series (one-hour cadence). Time-series for the four mean $P_a$ frequency intervals, the $\Phi_m$ and $I_c$ values (6 physical quantities), for each one of the 63 tiles of the 9-by-9 grid in Figure~\ref{fig:Diagram} are used for training. The 9-by-9 grid has 81 tiles. The top and bottom rows are not used due to the geometric effect removal we apply to all data, bringing the total number of tiles to 63. This geometric effect normalization assumes that the top and bottom rows are parts of the quiet Sun and is further discussed by \cite{kasapis2024predicting}. The processed training dataset (ML-ready) has 17,010 timelines available (63~tiles $\times$ 40~ARs $\times$ 6~physical quantities), and it occupies no more than 150 MB of data storage per AR. 

\begin{figure}[H]
\begin{center}
\begin{algorithm}[H]
\caption{LSTM Training Process}
\label{alg:lstm_training}
\begin{algorithmic}[1]
    \STATE Set Hyperparameters and Initialize LSTM Model (Figure \ref{fig:Diagram}, top left)
    \STATE Load ML-Ready Dataset
    \STATE Remove Emerging/Non-Emerging Tile Imbalance
    \STATE Split in Training (Table~\ref{tab:Training}) and Testing (Table~\ref{tab:Testing}) 
    \STATE Initialize Optimizer 
    \FOR{Active Region \textbf{in} Training Dataset}
        \FOR{Tile \textbf{in} AR Grid}
            \STATE Split in Input and Target
            \STATE Timelines Processing (Sliding Window Approach)
            \STATE Move Data to GPU
            \STATE Set Scheduler and Learning Rate Decay
            \FOR{Current Epoch \textbf{in} Total Epochs Number}
                \STATE Forward Pass the Training Data through the LSTM Model to Obtain Output/Predictions
                \STATE Zero the Gradients of the Optimizer
                \STATE Calculate Loss Using Output and Targets
                \STATE Backpropagate Loss
                \STATE Take Gradient Step Using LR and Update Model Weights
                \STATE Perform Validation
            \ENDFOR
        \ENDFOR
    \ENDFOR
\end{algorithmic}
\end{algorithm}
\end{center}
\end{figure}

Similar to AR11158 (Figure \ref{fig:Diagram}, left), none of the 45 tracked ARs occupies more than 10 tiles out of the 81 present in the grid. Imbalanced datasets in machine learning are problematic because they can lead to biased models that are overfitted to the majority class, resulting in poor performance on the minority class, which, like in our application, is often the most critical to predict \citep{chen2024survey}. The imbalance between tiles that exhibit activity over time (`Emerging' tiles) and tiles that do not (`Quiet' or `Non-emerging' tiles) is dealt with by omitting the top and bottom three (predominantly non-active) rows of data during training. By not using part of the grid, we keep a balance between the number of active and non-active tiles. It was observed that without adding a weighting parameter, a model trained on the entire tracked patch tends to over-fit on the quiet Sun data. 

After removing the dataset's imbalance, the remaining 2430~timelines (9~tiles $\times$ 45~ARs $\times$ 6~physical quantities) are split into training and testing datasets as discussed in Section \ref{sec:Dataset}. The training of the models is performed by iterating through the different ARs and the tiles within them, updating the model's weights separately for each tile. This approach ensures that the model recognizes each tile sample as a distinct region of the Sun rather than interpreting tile batches as different feature types within the same area. As seen in Algorithm \ref{alg:lstm_training}, before the AR and Tile iterations begin, we initialize the optimizer and set the learning rate ($LR$) of our choice. In this work, we use the \texttt{Adams} optimizer provided by the \texttt{PyTorch} library\footnote{https://pytorch.org/}. For every AR (Algorithm \ref{alg:lstm_training}, line 6) and every Tile within the AR (line 7), the training data is split into inputs and targets (line 8). Here, the four different $P_a$ time series are grouped together with the equivalent $\Phi_m$ observation to create the input that will be used to predict the target. In this first part of our research, the target is set to be the $I_c$ timeline. 

The timelines are also processed using a sliding window approach. The number of input ($In$) and output ($Out$) data points, assuming that $In+Out<240$ (240 total observations for every AR), determines the size of the sliding window that moves through the 240-long timelines, creating multiple overlapping input-output data pairs. The number of input and output points are hyper-parameters set before the training process begins (Algorithm~\ref{alg:lstm_training}, line 1). Other hyper-parameters include the learning rate ($LR$), the number of layers ($LN$), the number of units ($U$), and the number of epochs ($E$), as presented in Table~\ref{tab:Models}. This table lists the hyper-parameter values for the ten best-performing ML models trained within the parameter space explored in this research. For example, the best performing model (M8, marked in bold on Table~\ref{tab:Models}) has an input of 110 observations, corresponding to 110 hours (1 hour cadence), and an output of 12 observations, therefore a 122 hours wide window. This window is slid through the 240-hour timeline 118 times ($240-(In+Out)$), generating 118 distinct input-output pairs.

Before the training iterations begin (Algorithm~\ref{alg:lstm_training}, line 12) and the model's weights start being updated, the scheduler is initialized using the \texttt{StepLR} PyTorch function. Subsequently, the training data is transferred to the Graphics Processing Unit (GPU) to enable the network’s forward pass and weight update computations to be executed there. We use the CUDA parallel computing platform available through PyTorch to train models on the NASA High-End Computing (HECC\footnote{https://www.nas.nasa.gov/hecc/}) GPUs and make the training process more time-efficient. Models with layers, units, and epoch numbers similar to those of Model 8 (M8) take less than 2.5 hours to train. The scheduler for the models in Table~\ref{tab:Models} is configured to decrease the learning rate by 10\% with each new epoch. 

Within each epoch, a sequence of processes, common in ML, takes place. The input data (observed $P_a$ and $\Phi_m$) are passed through the LSTM model (Algorithm~\ref{alg:lstm_training}, line~13), and an output $I_c$ prediction vector, matching in size ($Out$) the target vector, is produced. The optimizer gradients are zeroed (line 14) for the new ones to be calculated during backpropagation (line 16) based on the new loss calculated in line 15. In this research, the loss is determined using the Mean Square Error (MSE) formula provided by the \texttt{torch.nn.MSELoss()} PyTorch function by comparing the output of each forward pass to the true observations. The model weights get updated using the calculated gradients and the learning rate $LR$ (line 17), while validation is also performed during every iteration to ensure the proper loss reduction. 

\begin{table}[htbp]
\label{tab:Models}
\begin{ruledtabular} 
\begin{tabular}{ccccccccccccc}
\multicolumn{1}{c}{} & \multicolumn{6}{c}{\textbf{Model Parameters}} & \multicolumn{6}{c}{\textbf{Average Testing RMSE for 12h Prediction}} \\
 \cline{2-7} \cline{8-12}
\textbf{Model} & In & Out & LN & U & E & LR & AR11698 & AR11726 & AR13165 & AR13179 & AR13183 & \textbf{Avg} \\ \hline
M1 & 110  &  5  &  3  & 64  & 1000  & 0.01  & 0.106  & 0.084  & 0.107  & 0.140  & 0.115  & 0.110 \\ \hline
M2 & 110  &  8  &  3  & 64  & 1000  & 0.01  & 0.130  & 0.092  & 0.113  & 0.145  & 0.107 & 0.117 \\ \hline
M3 & 120  &  12  &  3  & 32 & 1000  & 0.001  &  0.118  & 0.111  & 0.122 & 0.160  & 0.115  & 0.125 \\ \hline
M4 & 120  &  12  &  3  & 64 & 1000 & 0.001  & 0.125  & 0.106  & 0.128  & 0.150  & 0.118  & 0.125 \\ \hline
M5 & 120  &  12  &  3  & 64  & 1000  & 0.0005  & 0.121  & 0.105  & 0.121  & 0.161  & 0.116  & 0.125 \\ \hline
M6 & 120  &  12  &  4  & 64 & 1500  & 0.0005  & 0.126  & 0.105  & 0.121  & 0.152  & 0.144  & 0.130 \\ \hline
M7 & 110  &  12  &  3  & 128 & 1000 & 0.01  &  0.124  &  0.090  &  0.111  & 0.125  &  0.100 & 0.110 \\ \hline
\textbf{M8} & \textbf{110}  &  \textbf{12}  &  \textbf{3}  & \textbf{64}  & \textbf{1000}  & \textbf{0.01}  & \textbf{0.128} & \textbf{0.108}  &  \textbf{0.127} & \textbf{0.158}  & \textbf{0.125}  & \textbf{0.130} \\ \hline
M9 & 110  &  18  &  3  & 64 & 1000 & 0.01  & 0.136 & 0.099 & 0.125  & 0.181  & 0.122  & 0.133 \\ \hline
M10 & 110  &  24  &  3  & 64 & 1000 & 0.01  & 0.136  & 0.100  & 0.128  & 0.153 & 0.136  & 0.131 \\ \hline
 &   &    &   &  & AR & Average & 0.125  & 0.101  & 0.120  & 0.153 & 0.120  & 0.124 \\ \hline
\end{tabular}
\end{ruledtabular}
\caption{Summary of the 10 best performing models we trained during this research with their defined hyperparameters and the average, across tiles, testing RMSE for a 12h prediction setup. Testing was performed on the five ARs of Table~\ref{tab:Testing}.}
\end{table}

In this research, an LSTM Encoder-Decoder architecture is selected as demonstrated in the top right diagram of Figure~\ref{fig:Diagram}. The network architecture depends on two hyperparameters: the number of layers $LN$ and the number of units within each layer $U$. More specifically, the models presented in Table~\ref{tab:Models} consist of an LSTM Encoder which encodes the input and forwards the cell state ($c$) and the hidden state ($h$) to an LSTM Decoder, to decode the information and, with the help of a fully connected (FC) layer, produce a prediction $y$. For each model, both the encoder and the decoder contain $LN$ LSTM layers (\texttt{nn.LSTM} class in PyTorch) and each layer contains $U$ LSTM units. More than 200~models were trained, exploring parameter spaces for different combinations of hyperparameters. The models that, when tested, produced average RMSE values $\leq 0.133$ across all five testing ARs, are presented in Table~\ref{tab:Models}.

As soon as training is over, testing takes place after freezing (preventing them from changing) the weights of the model and by forward passing the time-series data from the 5 testing ARs. Here, the $I_c$ time series, rather than being used as targets for training, are used as the observed true values to which we compare the prediction outputs and evaluate the models. The $E$ and $LR$ hyperparameters do not matter during testing, the $LN$ and $U$ remain the same as they are related to the structure of the model, therefore $In$ and $Out$ are the only variables that can differ between training and testing. For instance, even if a model was trained with $Out_{train}=5$, it can be tasked during testing to predict 24 observations ahead ($Out_{test}=24$), instead of 5. Similarly, although a model might be trained by inputting 120 observations ($In_{train}=120$), we can test it by inputting a smaller number of observations. This is useful in cases like AR11726 (Table~\ref{tab:Models}), where the emergence of the AR (the stage of the AR evolution that we care the most about) happens closer to the East limb. In such cases, less observed data before emergence are available, and therefore, prediction needs to happen with smaller $In$ values. To avoid any confusion, in this research, the output vector is always 12 predictions long ($Out_{test} = 12$), from which we use the very last element, as it represents the 12-hour-ahead prediction. Similarly, the input vector is always 96 hours ($In_{test} = 96$), except for AR11726, for  which it is 72 hours for the reasons discussed above.

Testing of the model performance is done in two ways: a) by calculating the Mean Squared Error (RMSE) between the predicted and the observed $I_c$ time series and b) by comparing the timestamps of true and predicted emergence (the entirety of Section \ref{sec:Results} is dedicated to this method). For the 12 hours prediction problem discussed here, and given that we input observations taken over 96 hours from a total of 240 observations, we are able to do predictions for $n = 132$ hours. This 132 entries-long prediction vector $y_{pred}$ is subtracted by the corresponding vector of observed values $y_{obs}$ for the RMSE to be calculated using Equation~\ref{eq:RMSE}: 

\begin{equation}
\label{eq:RMSE}
\text{RMSE} = \sqrt{\frac{1}{n} \sum_{i=1}^{n} \left( y_{obs} - y_{pred} \right)^2},
\end{equation}
where $n$ is the number of observations, and $y$ is the observed and predicted $I_c$ values.

\begin{figure}[ht]
\centering
\includegraphics[width=\textwidth]{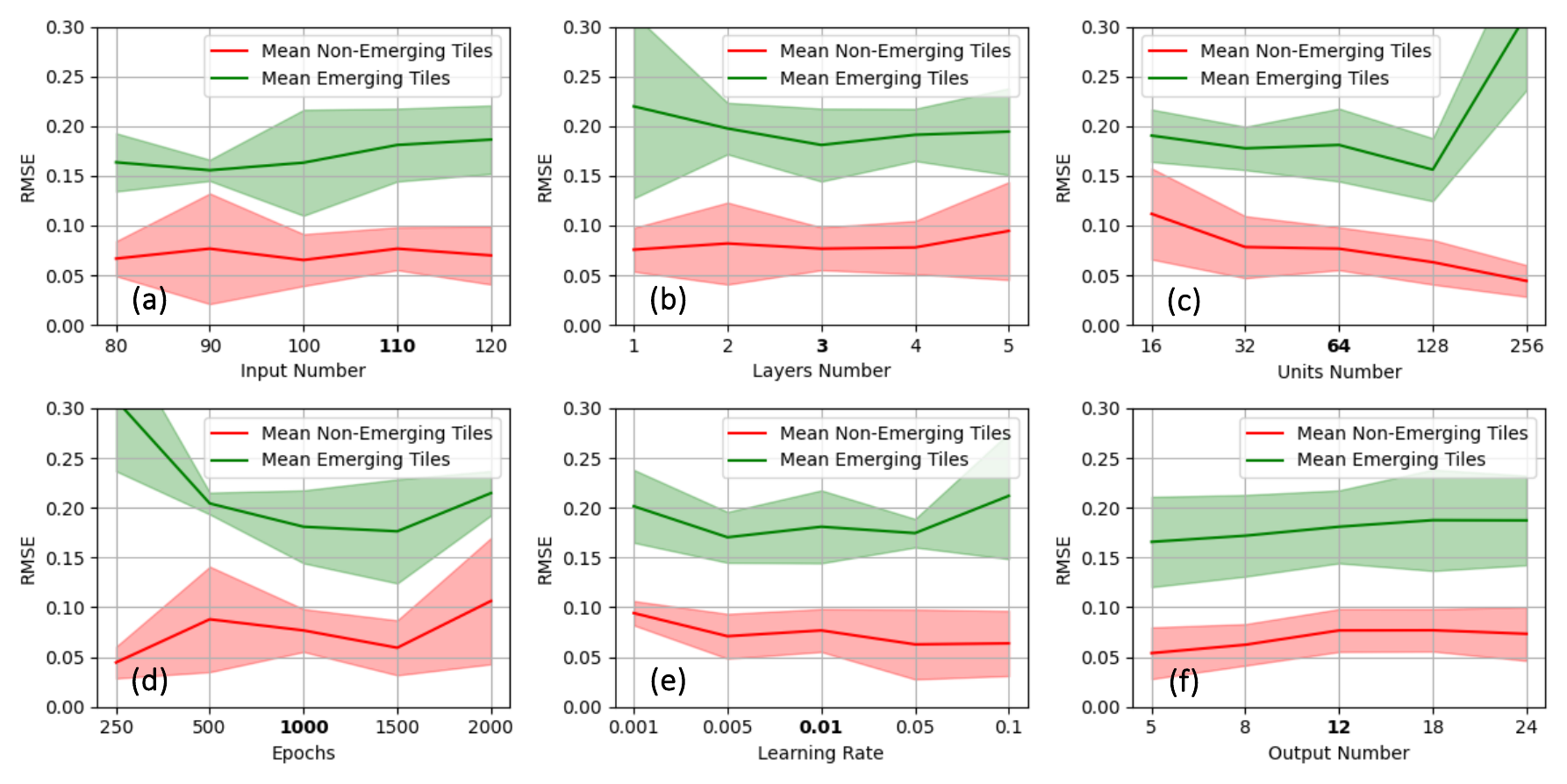}
\caption{RMSE obtained during the testing of Model 8 variations which differ in the number of inputs (panel a), layers (b), units (c), epochs (d), outputs (f), and the leaning rate (e). The green lines represent the mean RMSE for the tiles in which emergence has been observed, whereas the red lines represent the mean RMSE for the non-emerging tiles. The red and green shadows represent the standard deviation. All models were evaluated on the standard 12-hour prediction problem using 96 hours worth of input data.}
\label{fig:hyperparameters} 
\end{figure}

Table~\ref{tab:Models} presents the average RMSE across all tiles that were tested (four emerging and three non-emerging tiles) using the five testing ARs. The average RMSE value across all models for AR11726 is 0.101, compared to 0.153 for AR13179. Althought this difference shows that it is more difficult for all models to accurately predict the variations of $I_c$ in AR13179 compared to AR11726, it does not necessarily mean that predicting the emergence of AR13179 is more difficult than predicting that of AR11726. This is because the RMSE measures the agreement between predicted and observed time series, but it does not include any information about the success of the AR emergence prediction. Regardless, the RMSE is a good first metric for evaluation of model performance during the phase of hyperparameter space exploration. The spaces that produced low RMSE were explored again by varying one of the hyperparameters each time. An example of a space that was explored is the one of Model~8 (M8; Figure~\ref{fig:hyperparameters}).

Model 8 has 3 layers, each one comprised of 64 units, in its encoder and decoder, and was trained using $In = 110$ hours, $Out = 12$ hours, $E = 1000$ iterations, and $LR = 0.01$. In panel (a) of Figure \ref{fig:hyperparameters} we present the RMSE results of the four Model 8 variations that have $In = 80, 90, 100, 120$ instead of 110 that Model 8 originally has. Panel (b) presents results for another four Model 8 variations trained using the original $In = 110$, but with $LN = 1, 2, 4, 5$ compared to 3 layers that Model 8 originally has. Similarly, panels (c), (d), (e), and (f) present variations of Model 8 with variable numbers of $U$, $E$, $LR$, and $Out$, respectively. In all six plots of Figure \ref{fig:hyperparameters}, the RMSE results of Model 8 are marked in bolt. All 25 models (24 variations and Model 8) presented, although trained using different hyperparameters, are tested using 96-minute inputs to predict the $I_c$ 12 hours ahead. This is the testing setup used for all results presented in this paper. 

As mentioned, within the tracked solar regions included in the dataset, there are some tiles that experience emergence of magnetic flux (emerging tiles, Figure~\ref{fig:hyperparameters}) and some that never experience activity (non-emerging tiles) during the 240 hours of tracked observations. While Table~\ref{tab:Models} reports the average RMSE values for all tiles, Figure \ref{fig:hyperparameters} distinguishes between the average RMSE for non-emerging tiles (red) and emerging tiles (green), presenting them separately. In all plots, the RMSE values of the non-emerging tiles are always smaller than those of the emerging tiles, which shows that Model 8 and its variations are very good at predicting that non-emerging regions will remain quiet. The shadow behind each RMSE line indicates the standard deviation between the scores of the five ARs used for the model testing. A more extended shadow behind a model's RMSE value means that the model had difficulty predicting some ARs compared to others. Models with smaller RMSE standard deviations (e.g., Model 8) tend to be more reliable in reconstructing the evolution of continuum intensity.

In the search for the model that performs the best, when only taking into consideration the RMSE, one could conclude by observing Figure~\ref{fig:hyperparameters}, that Model 8 uses the number of layers that produces the least erroneous results but could benefit from a slight increase in the number of epochs and units. There are two major drawbacks we need to consider when increasing the number of epochs, $E$, and units, $U$. The most obvious is the training time. A 128-unit model takes almost double the time to train compared to a 64-unit model, in the same way that a 1500-epoch training would increase the training time by 50\%. The second and most important reason for picking Model~8 over the seemingly better (in terms of RMSE) variants is that it provides more accurate alarms for the emergence prediction problem than the other models. The way we define the emergence of an active region and the way we evaluate the models based on such a definition will be the main topic of discussion in the next Section. 


\section{Active Region Emergence Prediction} \label{sec:Results}

There is currently no community-wide accepted definition of active region emergence. For instance, NOAA defines emergence every 24 hours, as they issue a Solar Region Summary\footnote{https://www.swpc.noaa.gov/products/solar-region-summary} report at the end of each UTC day. Sometimes, the time of AR emergence is defined as the time when the emerging magnetic flux is associated with strong diverging flows and decreased continuum intensity following the formation of the active region. To be able to evaluate the performance of the developed ML models, we define AR emergence as the time when continuum intensity decreases for more than 3 hours with speed over 0.01 (in relative units):

\begin{equation} \label{eq:criterion}
\dfrac{dI_c}{dt} < -0.01 \quad \text{for} \quad t_{sus} > 3 \text{ hours},
\end{equation}
where $I_c$ is continuum intensity (predicted or observed), and $t_{sus}$ is the sustained time. This definition of a moment of AR emergence reflects what an observer would define as emergence during visual inspection.

\begin{figure}[ht]
\centering
\includegraphics[width=\textwidth]{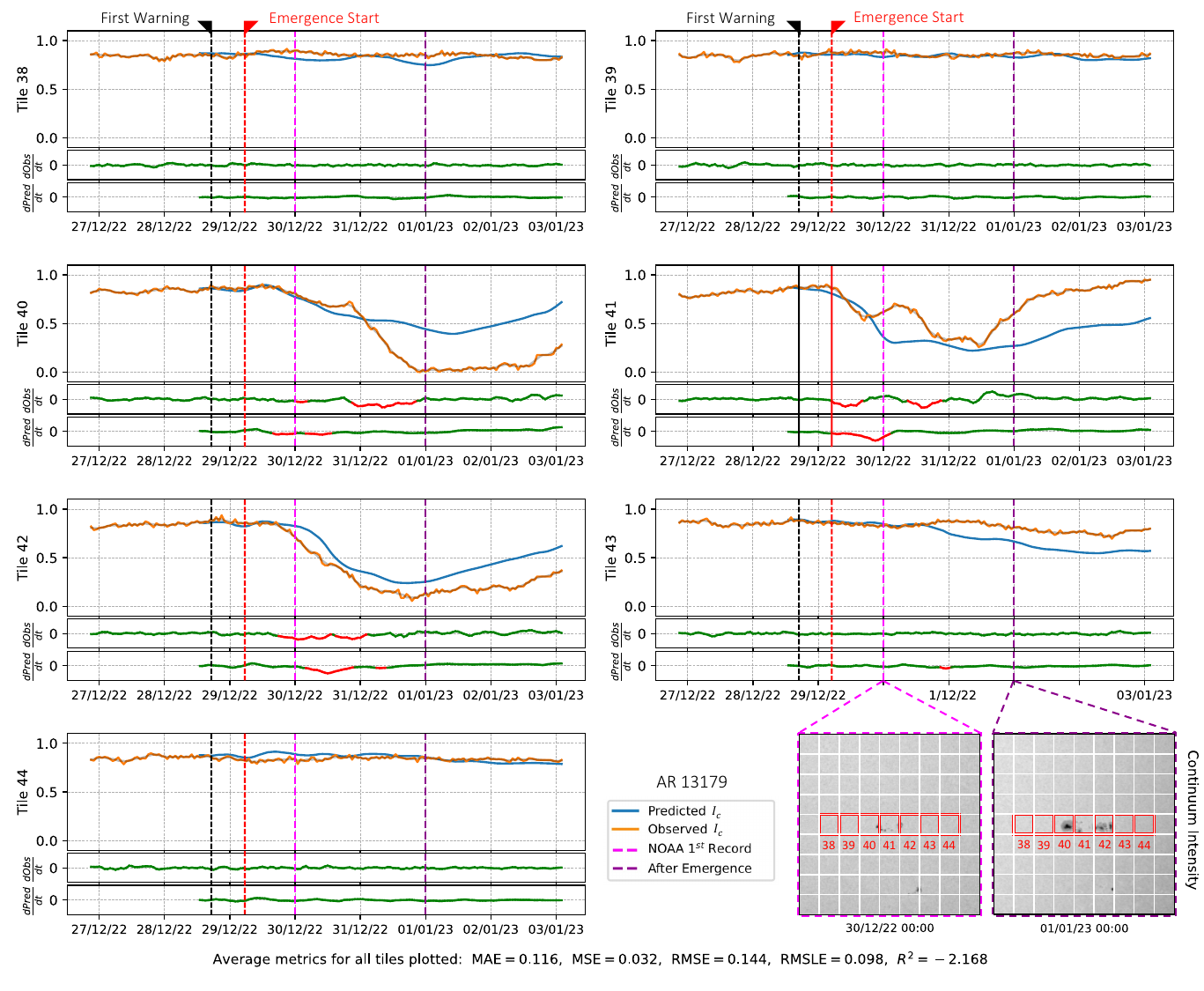}
\caption{Evaluation of Model~8 on predicting variations of the $I_c$ for selected tiles of AR13179 (tiles 38-44). The tiles' locations are marked in red squares at the bottom right continuum intensity images. Each tile's corresponding observed and predicted mean continuum intensity variations are shown as orange and blue curves. The time-derivatives of the continuum intensity (observed and predicted) are shown color-coded according to the Equation~\ref{eq:criterion} criteria. In red are the emergence periods while in green are the non-emerging (quiet) states. Tiles 40-42 are `active' because a decrease in the continuum intensity is observed, while tile 41 exhibits the first signatures of AR13179 emergence. The rest of the tiles are quiet and correspond to non-emerging states. Vertical dashed lines identify the following moments: NOAA's first record of the active region (magenta), the time two days after NOAA's first record (purple), the time when Model~8 produces its first emergence alarm (First Warning, black) and the time when the observed emergence starts (Emergence Start, red). The tiles for which instead of dashed, the First Warning and Emergence lines are solid, are the tiles in which these events took place.} 
\label{fig:results13179}
\end{figure}

For instance, we evaluate Model~8 (Table~\ref{tab:Models}) on predicting the emergence of AR13179. Each panel in Figure~\ref{fig:results13179} represents the observed (orange curves) and predicted (blue curves) evolution of the continuum intensity for the seven tiles (38 - 44) enclosed in red squares in the bottom right $I_c$ maps. The time derivative for observed and predicted continuum intensity, $dI_c/dt$, is included at the bottom of each panel. Color-coded in green is the non-emergence state, while red is emergence - the parts of the $dI_c/dt$ curves that fulfill Equation~\ref{eq:criterion}. The LSTM model can distinguish between tiles that remain quiet throughout observations and tiles that exhibit activity. For example, the model predictions show a good agreement with observations in both cases when emergence is apparent (e.g., tiles 40-41; Figure~\ref{fig:results13179}), and when the analyzed areas remain quiet (e.g., tiles 38-39). 

Typically, the model prediction deviates from observations after AR emergence (e.g., tiles 42-43). This discrepancy can be explained by the interaction of emerging magnetic fields and the reorganization of the convection structure at the photosphere, which affects the properties of the power maps. Another reason for the appearance of such discrepancies is that the training dataset is mainly focused on the period during which the AR is emerging, rather than the later AR evolution period. Despite these limitations, we can often see a remarkable qualitative agreement with observations over significant time of the AR evolution (e.g., tiles 40-42 for AR13179; Figure~\ref{fig:results13179}).

To identify the moment of AR emergence both in predictions and observations, we use criteria that describe the decreased rate of the continuum intensity (Equation~\ref{eq:criterion}). During the emergence and formation of the active region, the magnetic flux can spread across several tiles. Therefore, we assume that the AR emergence start time (predicted or observed) is the earliest moment when the emergence criterion is satisfied (e.g., tile 41 in Figure~\ref{fig:results13179}). Thus, the model predicted emergence on 2022-12-28 18:00 (black dotted line in Figure~\ref{fig:results13179}, First Warning), half a day before the observed emergence at 2022-12-28 06:00 (red dotted line in Figure~\ref{fig:results13179}, Emergence Start). The satisfaction of this criterion for other tiles (red $dI_c/dt$) is caused by the expansion of the emergence of these tiles or the movement of magnetic flux together with diverging flows during emergence.

While it is logical for the first tile where emergence is observed to also be the one where Model 8 triggers its initial alarm, it is not always the case, as $P_a$ flows move and can potentially trigger the Model 8 emergence alarm in neighboring tiles. One such case is AR11726 (Figure \ref{fig:results11726} in Appendix), where although the first activity warning is produced in Tile 41 at 2013-04-19 14:00, the first observed emergence happens on the neighboring Tile 42 at 2013-04-19 19:00. The existence of incorrect local activity predictions (ILAP) can also be attributed to this movement of flows between the arbitrarily chosen tiles of our analysis and they will be further discussed later in this text. Nonetheless, for four out of the five ARs we have included in our testing dataset, our model signals its first activity alarm (moment of the emergence prediction) not only at the same tile it was observed, but also many hours before it was observed, as seen in the AR Emergence Prediction (Experiment) column of Table \ref{tab:forecasts}.

If we consider tiles as independent instances, we can evaluate the local continuous intensity evolution and the possibility of predicting the magnetic flux increase. This approach can be useful for future studies to capture situations when emergence takes place in different tiles close to each other. Tile~42 for example (Figure~\ref{fig:results13179}), based on the observed $I_c$, presents its first observed activity at 2022-12-29 19:00 while the first activity is predicted 12 hours ahead for 2022-12-30 05:00, therefore Model~8 produces its first activity alarm at 2022-12-29 17:00, 2 hours ahead of the observed activity. Similarly, tile 40 has its first activity observed at 2022-12-30 01:00 while the first activity prediction is at 2022-12-29 17:00, which means that the first activity alarm was produced at 2022-12-29 05:00, 20 hours ahead of the observed. Therefore, if we treat each tile of AR13179 independently, activity on tile 40 is predicted 20 hours ahead, on tile 41, 12 hours ahead, and on tile 42, 2 hours ahead. These independent tile AR emergence forecast times are also tabulated on Table \ref{tab:forecasts}, under AR13179 and the respective tile number (Activity Prediction for Selected Tiles columns). Following the same analysis for the rest active regions used for testing (AR11698, AR11726, AR13165, and AR13183) we obtain the remaining forecast times in Table~\ref{tab:forecasts}. Corresponding plots for these testing ARs, can be found in the Appendix (Figures~\ref{fig:results11698} -- \ref{fig:results13183}). 

All models presented, during testing, had output vectors of length $Out = 12$ hours (prediction window), with a data cadence of 1 hour. In reality, there are two reasons for which, in an operational setting, where predictions are produced in real-time, this model would only be able to predict values 5 hours in advance. The first reason stems from the definition of the AR emergence given in Equation~\ref{eq:criterion}, which requires the decrease of the derivative under a specified threshold for 3 consecutive hours. In an operational setting, to produce an alarm, you would have to wait for 3 hours until the Equation \ref{eq:criterion} condition gets fulfilled, and therefore 3 hours after the first alarm noted in Figures~\ref{fig:results13179} -- \ref{fig:results13183}. Another four hours are lost due to the process of calculating $P_a$ maps. To calculate the $P_a$ maps we use 8-hour Dopplergram sequences, but here the recorded power map time is not the last time-stamp of the sequence, but the midpoint. For this reason, in an operational setting, the entire analysis would have to be shifted by four hours. Accounting for the 3 hours lost because of the emergence definition and another 4 hours because of the Dopplergram time definition, a total of 7 hours of prediction window is lost when simulating an operational environment.

\begin{table}[t] \label{tab:forecasts}
\centering
\begin{tabular}{cccccccccl}
  & \multicolumn{7}{c}{Activity Prediction for Selected Tiles}   & \multicolumn{2}{c}{AR Emergence Prediction}  \\ \hline
\multicolumn{1}{|c|}{Tile Type}  & Quiet  & Quiet   & Quiet  & Q/A & Active   & Active    & \multicolumn{1}{c|}{Active}  & \multicolumn{1}{c|}{Experiment}  & \multicolumn{1}{l|}{Operational}   \\ \hline
\multicolumn{1}{|c|}{AR11698}   & Tile 47 & Tile 48  & Tile 50  & Tile 53  & {\bf Tile 49}  & Tile 51  & \multicolumn{1}{c|}{Tile 52}   & \multicolumn{2}{c|}{Tiles 49/49}    \\
\multicolumn{1}{|c|}{Figure \ref{fig:results11698}} & {\color[HTML]{009901} Quiet} & {\color[HTML]{009901} Quiet} & {\color[HTML]{CB0000} ILAP}    & {\color[HTML]{009901} Quiet}      & {\color[HTML]{FF7F00} {\bf 4h Alarm}}  & {\color[HTML]{009901} 12h Alarm} & \multicolumn{1}{c|}{{\color[HTML]{FF7F00} 1h Alarm}}  & {\color[HTML]{FF7F00} 4h Alarm}  & \multicolumn{1}{l|}{{\color[HTML]{CB0000} -3h Alarm}} \\ \hline
\multicolumn{1}{|c|}{AR11726}   & Tile 38  & Tile 39  & Tile 40   & Tile 41 & {\bf Tile 42}  & Tile 43 & \multicolumn{1}{c|}{Tile 44}          & \multicolumn{2}{c|}{Tiles 42/41}  \\
\multicolumn{1}{|c|}{Figure \ref{fig:results11726}} & {\color[HTML]{009901} Quiet} & {\color[HTML]{009901} Quiet} & {\color[HTML]{009901} Quiet} & {\color[HTML]{009901} 39 h Alarm} & {\color[HTML]{009901} {\bf 9h Alarm}}  & {\color[HTML]{009901} 39h Alarm} & \multicolumn{1}{c|}{{\color[HTML]{009901} 18h Alarm}} & {\color[HTML]{009901} 17h Alarm}  & \multicolumn{1}{l|}{{\color[HTML]{009901} 10h Alarm}} \\ \hline
\multicolumn{1}{|c|}{AR13165} & Tile 29 & Tile 30 & Tile 35 & Tile 31 & {\bf Tile 32} & Tile 33  & \multicolumn{1}{c|}{Tile 34}   & \multicolumn{2}{c|}{Tiles 32/32}                                                         \\
\multicolumn{1}{|c|}{Figure \ref{fig:results13165}} & {\color[HTML]{009901} Quiet} & {\color[HTML]{009901} Quiet} & {\color[HTML]{009901} Quiet} & {\color[HTML]{009901} 35h Alarm}  & {\color[HTML]{009901} {\bf 36h Alarm}} & {\color[HTML]{FF7F00} 7h Alarm}  & \multicolumn{1}{c|}{{\color[HTML]{009901} 55h Alarm}} & {\color[HTML]{009901} 36h Alarm} & \multicolumn{1}{l|}{{\color[HTML]{009901} 29h Alarm}} \\ \hline
\multicolumn{1}{|c|}{AR13179} & Tile 38 & Tile 39 & Tile 43 & Tile 44 & Tile 40 & Tile 41  & \multicolumn{1}{c|}{\textbf{Tile 42}}  & \multicolumn{2}{c|}{Tiles 41/41}  \\
\multicolumn{1}{|c|}{Figure \ref{fig:results13179}} & {\color[HTML]{009901} Quiet} & {\color[HTML]{009901} Quiet} & {\color[HTML]{CB0000} ILAP}    & {\color[HTML]{009901} Quiet}      & {\color[HTML]{009901} 20h Alarm} & {\color[HTML]{009901} {\bf 12h Alarm}} & \multicolumn{1}{c|}{{\color[HTML]{FF7F00} 2h Alarm}}  & {\color[HTML]{009901} 12h Alarm} & \multicolumn{1}{l|}{{\color[HTML]{009901} 5h Alarm}}  \\ \hline
\multicolumn{1}{|c|}{AR13183}  & Tile 38 & Tile 39 & Tile 44 & Tile 40 & {\bf Tile 41} & Tile 42 & \multicolumn{1}{c|}{Tile 43}  & \multicolumn{2}{c|}{Tiles 41/41}  \\
\multicolumn{1}{|c|}{Figure \ref{fig:results13183}} & {\color[HTML]{009901} Quiet} & {\color[HTML]{009901} Quiet} & {\color[HTML]{009901} Quiet} & {\color[HTML]{CB0000} -2h Alarm}  & {\color[HTML]{FF7F00} {\bf 5h Alarm}}  & {\color[HTML]{FF7F00} 4h Alarm}  & \multicolumn{1}{c|}{{\color[HTML]{009901} 11h Alarm}} & {\color[HTML]{FF7F00} 5h Alarm}  & \multicolumn{1}{l|}{{\color[HTML]{CB0000} -2h Alarm}} \\ \hline
\end{tabular}
\caption{Summary table of activity prediction for independently analyzed tiles and AR emergence prediction results. The tiles tested in each AR are categorized in Quiet (non-emerging) and Active (Emerging). Tiles where the emergence of the active region is observed are marked with bold fonts. The alarms that signified the emergence of an AR (first alarm within AR) are also marked in bold. In red are the two incorrect local activity predictions (ILAP) and the delayed activity alarms. In orange are the activity and emergence alarms that in an operational setting would be delayed ($\leq7$ hours) while in green are the correct alarms.}
\end{table}

In Table~\ref{tab:forecasts}, we summarize results for selected tiles of five $30.66\times30.66$ degrees areas. Some of these tiles exhibit an increase in magnetic flux, and some of them remain quiet. For all testing ARs, presented in Table~\ref{tab:forecasts} are the results for 4 emerging (Active) and 3 non-emerging (Quiet) tiles, except for AR11698 and AR13179, for which only three tiles present activity. The prediction values were produced assuming an experimental setting where the entire 12-hour forecast window is available and the operational constraints (7 hour prediction delay) were not applied. The values in orange are the cases where, although Model~8 predicted the tile activity or AR emergence experimentally, the result would be a late alarm if the operational constraints were to be applied. 

The operational predictions for the entire ARs are tabulated at the right-most column in Table~\ref{tab:forecasts}. These values represent the experimental forecast values with the 7-hour delay (due to operational processing) subtracted. Model 8 is able to predict successfully and in an operational setting AR11726, AR13165, and AR13179, 10, 29, and 5 hours ahead respectively (AR Emergence Prediction columns). Note that for AR13179, Model~8 produced its first alarm in a different (neighboring) tile than the one in which emergence was observed. Late operational alarms were produced for AR11698 and AR13183, but only for 3 and 2 hours, respectively. Independent LSTM analysis for selected tiles of the five $30.66\times30.66$ degrees solar disc areas shows correct prediction for 26 out of the 35 tiles (74\%) in an operational setting. The prediction success rate is higher for quiet tiles, with only 2 ILAPs and 15 correct true negatives (88\%), whereas for the active tiles, there are 11 operationally correct forecasts and 7 late alarms (61\%).

\section{Incorrect Local Activity Predictions} \label{sec:IAPs}

Treating each tile independently allows us to explore the sensitivity of the LSTM model to mean evolution variations of the convection oscillatory properties, the continuum intensity, and present changes of magnetic flux. It is important to note that during AR emergence, a strong diverging flow may transport magnetic flux from one tile to another, which may cause strong variations in the acoustic power. Therefore, after the first alarm of AR emergence (tile 41, Figure~\ref{fig:results13179}), the LSTM model correctly predicts the incoming activity of neighborhood tiles (tiles 40 and 42). In this section we are discussing a few cases when LSTM model incorrectly predicted such activity.

\subsection{AR11698, Tile 50}

\begin{figure}[h]
\label{fig:FAs}
\centering
\includegraphics[width=\textwidth]{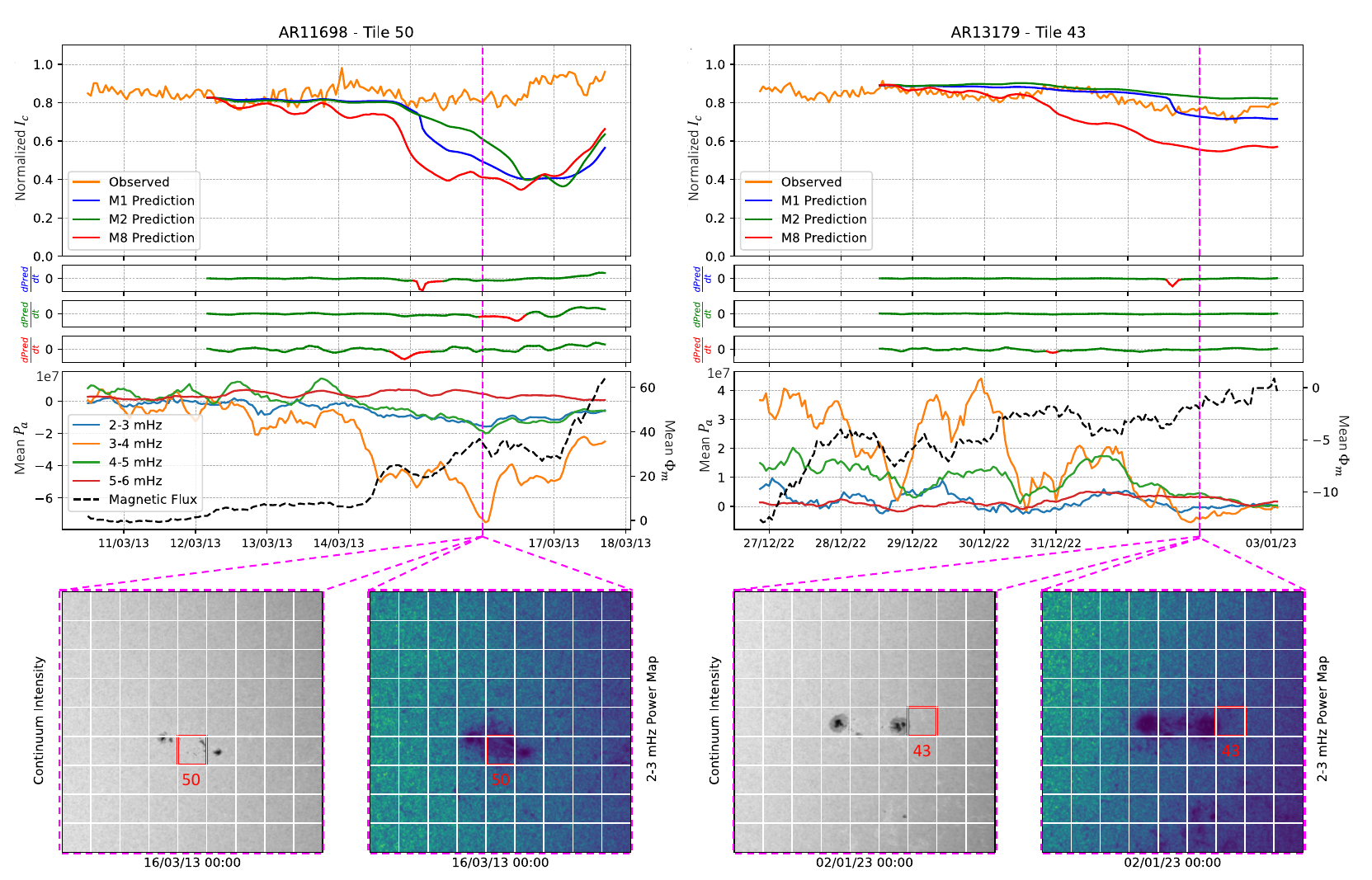}
\caption{Comparison of the continuum intensity predictions obtained using Models 1, 2, and 8 (upper plots), the corresponding intensity derivatives ($dPred/dt$) color-coded according to the Equation \ref{eq:criterion} criterion (plots in the middle), and the variations of mean acoustic power and unsigned magnetic flux (plots in the bottom) for two ILAP cases near AR11698 (left) and AR13179 (right). These selected tiles marked in red squares over the continuum intensity images and 2-3~mHz power maps on the bottom of the figure, corresponding to about two days after emergence.}
\label{fig:false_alarms}
\end{figure}

The LSTM activity prediction in the vicinity of AR11698 (Figure \ref{fig:results11698}) shows incorrect local activity predictions (ILAP) in tile 50 (red part of derivative curves in Figures~\ref{fig:results11698} and \ref{fig:false_alarms}). The decrease of the continuum intensity predicted by different LSTM models (Figure~\ref{fig:false_alarms}, upper left panel) was driven by significant suppression of the 3-4~mHz mean power that precedes the increase in the unsigned magnetic flux for 4 hours, while other frequency ranges don't show strong changes with time. Interestingly, the Model 8 is more sensitive to changes in the integrated properties of oscillations, while models Models 1 and 2 reveal less sensitivity to the background conditions at the photosphere. 

The origin of the discrepancies between predicted and observed continuum intensity is tightly connected with the magnetic flux evolution during the emergence and formation of AR11698. In this case, the emergence of the active region was initiated in tile 49. During its emergence and evolution, the two-polarity magnetic flux quickly moved in the opposite direction, forming AR11698. The tile 50 is located between opposite-polarity sunspots. Therefore, a mixture of magnetic patches of both polarities from one side exhibits elevation of unsigned magnetic flux, which causes suppression of the power of stochastic oscillations and, at the same time, prevents them from organizing magnetic fields into a coherent structure such as sunspots or pores, keeping the continuum intensity similar to a quiet region.

\subsection{AR13179, Tile 43}

Similarly, Models 1 and 8 predict some magnetic activity on tile 43 in the vicinity of emerging AR13179 (Figures~\ref{fig:results13179} and~\ref{fig:false_alarms}, right upper panels) while when observed, no significant changes in the mean continuum intensity are detected. In this case, even though emergence started in tile 42, magnetic flux shows a slow, gradual increase. In general, the mean power for the 3-4~mHz and 4-5~mHz frequency ranges show a significant enhancement and variations in time (Figure~\ref{fig:false_alarms}, right bottom plot), which makes interpretation of results challenging. These strong fluctuations in the power of oscillations were captured by Models 8 and 1 and were interpreted as a sign of upcoming activity. This example illustrates a possible way to improve our LSTM modeling, as, in its current version, the model considers relative variations of the acoustic power, which prevents it from recognizing the background enhancement in the acoustic power. Nevertheless, this existing limitation didn't prevent us from obtaining a reliable emergence prediction for AR13179.


\section{CONCLUSIONS AND FUTURE WORK} \label{sec:Conclusion}

This paper addresses the problem of predicting the emergence of active regions (ARs) by developing a dataset that includes 61 ARs tracked with solar rotation before and after emergence. This dataset was used to generate acoustic power time-series for four different frequency ranges. In this research only 45 ARs were utilized due to the presence of data gaps on the remaining 16 ARs. Using the acoustic power ($P_a$) and unsigned magnetic flux ($\Phi_m$) time series as input, we developed LSTM models to predict decreases in the continuum intensity ($I_c$) associated with the emergence of an AR. Despite utilizing four frequency ranges to predict AR emergence, we found that power maps for 3-4 and 4-5~mHz frequency ranges carry most of information related to coming emergence of an AR. 

The trained models' performance was tested on 5 ARs, unseen during training, using two evaluation criteria: the RMSE between predicted and observed intensity, and the comparison between true and predicted time of emergence, based on the Equation~\ref{eq:criterion} criterion. One of the best-performing models trained during this research, Model~8, succeeded in predicting the emergence of all 5 ARs in an experimental setting and 3 of them in an operational setting. The model predicted the emergence of AR11726, AR13165,  and AR13179, 10, 29, and 5 hours in advance, respectively, while variations of this model achieved average RMSE values that are as low as 0.06 for quiet (non-emerging) tiles and 0.16 for active (emerging) tiles. This LSTM network application demonstrates the ability of an ML model to capture solar $P_a$ anomalies and predict $I_c$ variations, resulting in the first trained network of its kind capable of forecasting AR emergence. By analyzing the inputs of the ML models during training and comparing them to the output predictions, we identify the necessary improvements in the model's training and data curation that will allow for better AR emergence predictions in the future.

The three models presented in Figure~\ref{fig:false_alarms} produce incorrect local activity predictions (ILAPs) because they are spatially agnostic. During training, there is no information provided to Models~1, 2 and 8 in regards to which AR each tile belongs to, which part of the AR it belongs, nor which are its neighboring tiles. This training scheme that assumes the independence of each tile does not allow the network to understand the relationship between $P_a$ variations in neighboring tiles. Because $P_a$ often flows between the arbitrarily chosen borders of the tiles, a decrease in $P_a$ on one tile can potentially lead to a decrease in $I_c$, not in the same tile but in a neighboring one, as seen in Figure~\ref{fig:false_alarms}. To solve this problem, methods that inform the model about the spatial arrangement of the tiles should be used. Spatially informed models will be able to relate the variations of the $P_a$ not only with the $I_c$ of the same tile but also $P_a$ and $I_c$ of different tiles. Assigning coordinates to the time series of each tile used as input during training could potentially lead to the reduction of ILAPs.

The analysis of the AR emergence results highlights potential improvements not only for the ML-ready dataset but also for the methods used to train and test the ML models. The 9-by-9 grid setup used here produces 81 tiles, for the majority of which ($>90\%$) no activity can be detected. This active-quiet tile imbalance is addressed by omitting the majority of quiet tile time series during training to create a balance between the two types of data. This training technique, although adequate for training the models presented, discards a large amount of training data, which can potentially carry useful information related to AR emergence. To address the imbalance between positive and negative instances but also train on all the data available, space weather prediction works \citep{kasapis2024forecasting} have either used weighting factors or have been randomly sampling different negative instances in every training epoch, two methods that are apt to this research too and could potentially improve the prediction capabilities of the models.

The prediction of ARs involves analyzing complex, temporal data collected from various parts of the solar surface, therefore capturing long-range dependencies between $P_a$, $\Phi_m$ and $I_c$ patterns is crucial. Although LSTM models are fit for such tasks, in recent years, Transformers \citep{vaswani2017attention} have been the most prominent alternative due to their self-attention mechanism, which allows them to simultaneously process different input data (ex., different $P_a$ frequency ranges) and establish relationships across long sequences more effectively than the sequential processing approach of LSTMs. This ability to maintain global context without losing information over long time spans is particularly beneficial for understanding the intricate and extended temporal patterns of emerging AR data. Furthermore, transformers’ parallel processing capability significantly accelerates the training of large datasets, which is essential if training is performed without discarding any non-emerging tiles in the training dataset, as in such cases, the training time increases at least by 8 times. 

Throughout this work, two metrics have been used to evaluate the LSTM models' capabilities of predicting the emergence of ARs: the RMSE metric and the emergence criterion. While RMSE is a commonly used criterion in ML time-series prediction, the emergence definition outlined in Equation~\ref{eq:criterion} was specifically devised for this research due to the absence of a precise AR emergence start time definition in the literature. This definition of the emergence time of an active region is a mathematical description of what a human, observing the $I_c$ timelines, would define as an AR: a sustained and substantial decrease of $I_c$. It should be understood that although not arbitrary, our definition is chosen to fit specifically the problem of predicting the 5 ARs in our testing dataset. Future works should revisit this definition by not only taking into account more than 5 AR samples, but also by considering more physical parameters (such as $\Phi_m$, $V_D$ etc.) in order to make it more generalized and applicable to different problems.

\section*{Acknowledgments}
We want to thank the NAS Visualization Team (Nina McCurdy, Timothy Sandstrom, Christopher Henze) for their help with this project’s visualizations. This work is supported by the NASA AI/ML HECC Expansion Program, NASA Heliophysics Supporting Research Program, and the NASA grants 80NSSC19K0630, 80NSSC19K0268, 80NSSC20K1870, and 80NSSC22M0162. The code and data used to produce the results presented in this manuscript can be found at \url{zenodo.com/tbd}


\bibliography{sample631}{}

\begin{thebibliography}{}
\expandafter\ifx\csname natexlab\endcsname\relax\def\natexlab#1{#1}\fi
\providecommand{\url}[1]{\href{#1}{#1}}
\providecommand{\dodoi}[1]{doi:~\href{http://doi.org/#1}{\nolinkurl{#1}}}
\providecommand{\doeprint}[1]{\href{http://ascl.net/#1}{\nolinkurl{http://ascl.net/#1}}}
\providecommand{\doarXiv}[1]{\href{https://arxiv.org/abs/#1}{\nolinkurl{https://arxiv.org/abs/#1}}}

\bibitem[{{Ali} {et~al.}(2024){Ali}, {Sadykov}, {Kosovichev}, {Kitiashvili},
  {Oria}, {Nita}, {Illarionov}, {O'Keefe}, {Francis}, {Chong}, {Kosovich}, \&
  {Marroquin}}]{ali2024predicting}
{Ali}, A., {Sadykov}, V., {Kosovichev}, A., {et~al.} 2024, \apjs, 270, 15,
  \dodoi{10.3847/1538-4365/ad0a6c}

\bibitem[{Atti{\'e} {et~al.}(2018)Atti{\'e}, Kirk, Thompson, Muglach, \&
  Norton}]{attie2018precursors}
Atti{\'e}, R., Kirk, M.~S., Thompson, B.~J., Muglach, K., \& Norton, A.~A.
  2018, Space Weather, 16, 1143

\bibitem[{Barnes {et~al.}(2014)Barnes, Birch, Leka, \&
  Braun}]{barnes2014helioseismology}
Barnes, G., Birch, A., Leka, K., \& Braun, D. 2014, The Astrophysical Journal,
  786, 19

\bibitem[{Birch {et~al.}(2012)Birch, Braun, Leka, Barnes, \&
  Javornik}]{birch2012helioseismology}
Birch, A., Braun, D., Leka, K., Barnes, G., \& Javornik, B. 2012, The
  Astrophysical Journal, 762, 131

\bibitem[{Chapman(1939)}]{chapman1939magnetic}
Chapman, S. 1939, Nature, 144, 266, \dodoi{10.1038/144266a0}

\bibitem[{Chen {et~al.}(2024)Chen, Yang, Yu, Shi, \& Chen}]{chen2024survey}
Chen, W., Yang, K., Yu, Z., Shi, Y., \& Chen, C. 2024, Artificial Intelligence
  Review, 57, 1

\bibitem[{Fischer \& Krauss(2018)}]{fischer2018deep}
Fischer, T., \& Krauss, C. 2018, European journal of operational research, 270,
  654

\bibitem[{Georgoulis(2013)}]{georgoulis2013toward}
Georgoulis, M.~K. 2013, Entropy, 15, 5022

\bibitem[{Gottschling {et~al.}(2021)Gottschling, Schunker, Birch, L{\"o}ptien,
  \& Gizon}]{gottschling2021evolution}
Gottschling, N., Schunker, H., Birch, A., L{\"o}ptien, B., \& Gizon, L. 2021,
  Astronomy \& Astrophysics, 652, A148

\bibitem[{Hartlep {et~al.}(2011)Hartlep, Kosovichev, Zhao, \&
  Mansour}]{hartlep2011signatures}
Hartlep, T., Kosovichev, A.~G., Zhao, J., \& Mansour, N.~N. 2011, Solar
  Physics, 268, 321

\bibitem[{Harvey {et~al.}(1996)Harvey, Hill, Hubbard, Kennedy, Leibacher,
  Pintar, Gilman, Noyes, Title, Toomre, {et~al.}}]{harvey1996global}
Harvey, J., Hill, F., Hubbard, R., {et~al.} 1996, Science, 272, 1284

\bibitem[{Ilonidis {et~al.}(2013)Ilonidis, Zhao, \&
  Hartlep}]{ilonidis2013helioseismic}
Ilonidis, S., Zhao, J., \& Hartlep, T. 2013, The Astrophysical Journal, 777,
  138

\bibitem[{Ilonidis {et~al.}(2011)Ilonidis, Zhao, \&
  Kosovichev}]{ilonidis2011detection}
Ilonidis, S., Zhao, J., \& Kosovichev, A. 2011, Science, 333, 993

\bibitem[{Jaeggli \& Norton(2016)}]{jaeggli2016magnetic}
Jaeggli, S.~A., \& Norton, A.~A. 2016, The Astrophysical Journal Letters, 820,
  L11

\bibitem[{{Jiao} {et~al.}(2020){Jiao}, {Sun}, {Wang}, {Manchester}, {Gombosi},
  {Hero}, \& {Chen}}]{claspflarepred}
{Jiao}, Z., {Sun}, H., {Wang}, X., {et~al.} 2020, Space Weather, 18, e02440,
  \dodoi{10.1029/2020SW002440}

\bibitem[{Kasapis {et~al.}(2024)Kasapis, Kitiashvili, Kosovich, Kosovichev,
  Sadykov, O'Keefe, \& Wang}]{kasapis2024forecasting}
Kasapis, S., Kitiashvili, I.~N., Kosovich, P., {et~al.} 2024, arXiv preprint
  arXiv:2403.02536

\bibitem[{{Kasapis} {et~al.}(2024){Kasapis}, {Kitiashvili}, {Kosovichev},
  {Stefan}, \& {Apte}}]{kasapis2024predicting}
{Kasapis}, S., {Kitiashvili}, I.~N., {Kosovichev}, A.~G., {Stefan}, J.~T., \&
  {Apte}, B. 2024, arXiv e-prints, arXiv:2402.08890,
  \dodoi{10.48550/arXiv.2402.08890}

\bibitem[{Korpi-Lagg {et~al.}(2022)Korpi-Lagg, Korpi-Lagg, Olspert, \&
  Truong}]{korpi2022solar}
Korpi-Lagg, M.~J., Korpi-Lagg, A., Olspert, N., \& Truong, H.-L. 2022,
  Astronomy \& Astrophysics, 665, A141

\bibitem[{Kosovichev {et~al.}(2001)Kosovichev, Duvall, \&
  Scherrer}]{kosovichev2001time}
Kosovichev, A., Duvall, T., \& Scherrer, P. 2001, Helioseismic Diagnostics of
  Solar Convection and Activity, 159

\bibitem[{Le {et~al.}(2019)Le, Ho, Lee, \& Jung}]{le2019application}
Le, X.-H., Ho, H.~V., Lee, G., \& Jung, S. 2019, Water, 11, 1387

\bibitem[{{Leka} {et~al.}(2013){Leka}, {Barnes}, {Birch}, {Gonzalez-Hernandez},
  {Dunn}, {Javornik}, \& {Braun}}]{leka2013helioseismology}
{Leka}, K.~D., {Barnes}, G., {Birch}, A.~C., {et~al.} 2013, \apj, 762, 130,
  \dodoi{10.1088/0004-637X/762/2/130}

\bibitem[{{Li} {et~al.}(2024){Li}, {Zheng}, {Li}, {Hou}, {Li}, {Zhang}, \&
  {Chen}}]{li2024survey}
{Li}, T., {Zheng}, Y., {Li}, X., {et~al.} 2024, \apj, 964, 159,
  \dodoi{10.3847/1538-4357/ad2e90}

\bibitem[{{Maharana} {et~al.}(2024){Maharana}, {Cramer}, {Samara}, {Scolini},
  {Raeder}, \& {Poedts}}]{maharana2024solar}
{Maharana}, A., {Cramer}, W.~D., {Samara}, E., {et~al.} 2024, Space Weather,
  22, e2023SW003715, \dodoi{10.1029/2023SW003715}

\bibitem[{McIntosh(1990)}]{mcintosh1990classification}
McIntosh, P.~S. 1990, Solar Physics, 125, 251

\bibitem[{Pesnell {et~al.}(2012)Pesnell, Thompson, \&
  Chamberlin}]{pesnell2012solar}
Pesnell, W.~D., Thompson, B.~J., \& Chamberlin, P. 2012, The solar dynamics
  observatory (SDO) (Springer)

\bibitem[{Rees-Crockford {et~al.}(2022)Rees-Crockford, Nelson, \&
  Mathioudakis}]{rees2022preemergence}
Rees-Crockford, T., Nelson, C., \& Mathioudakis, M. 2022, The Astrophysical
  Journal, 940, 109

\bibitem[{Scherrer {et~al.}(1995)Scherrer, Bogart, Bush, Hoeksema, Kosovichev,
  Schou, Rosenberg, Springer, Tarbell, Title, {et~al.}}]{scherrer1995solar}
Scherrer, P.~H., Bogart, R.~S., Bush, R., {et~al.} 1995, The soho mission, 129

\bibitem[{Scherrer {et~al.}(2012)Scherrer, Schou, Bush, Kosovichev, Bogart,
  Hoeksema, Liu, Duvall, Zhao, Title, {et~al.}}]{scherrer2012helioseismic}
Scherrer, P.~H., Schou, J., Bush, R., {et~al.} 2012, Solar Physics, 275, 207

\bibitem[{{Schunker} {et~al.}(2016){Schunker}, {Braun}, {Birch}, {Burston}, \&
  {Gizon}}]{schunker2016sdo}
{Schunker}, H., {Braun}, D.~C., {Birch}, A.~C., {Burston}, R.~B., \& {Gizon},
  L. 2016, \aap, 595, A107, \dodoi{10.1051/0004-6361/201628388}

\bibitem[{{Schunker} {et~al.}(2024){Schunker}, {Roland-Batty}, {Birch},
  {Braun}, {Cameron}, \& {Gizon}}]{schunker2024flux}
{Schunker}, H., {Roland-Batty}, W., {Birch}, A.~C., {et~al.} 2024, \mnras, 533,
  225, \dodoi{10.1093/mnras/stae1776}

\bibitem[{Sherstinsky(2020)}]{sherstinsky2020fundamentals}
Sherstinsky, A. 2020, Physica D: Nonlinear Phenomena, 404, 132306

\bibitem[{Singh {et~al.}(2016)Singh, Raichur, \& Brandenburg}]{singh2016high}
Singh, N.~K., Raichur, H., \& Brandenburg, A. 2016, The Astrophysical Journal,
  832, 120

\bibitem[{Stefan \& Kosovichev(2023)}]{stefan2023exploring}
Stefan, J.~T., \& Kosovichev, A.~G. 2023, The Astrophysical Journal, 948, 1

\bibitem[{Vaswani {et~al.}(2017)Vaswani, Shazeer, Parmar, Uszkoreit, Jones,
  Gomez, Kaiser, \& Polosukhin}]{vaswani2017attention}
Vaswani, A., Shazeer, N., Parmar, N., {et~al.} 2017, Advances in neural
  information processing systems, 30

\bibitem[{Waidele {et~al.}(2023)Waidele, Roth, Singh, \&
  K{\"a}pyl{\"a}}]{waidele2023strengthening}
Waidele, M., Roth, M., Singh, N., \& K{\"a}pyl{\"a}, P. 2023, Solar Physics,
  298, 30

\bibitem[{Zhang {et~al.}(2017)Zhang, Wang, Dong, Zhong, \&
  Sun}]{zhang2017prediction}
Zhang, Q., Wang, H., Dong, J., Zhong, G., \& Sun, X. 2017, IEEE geoscience and
  remote sensing letters, 14, 1745

\end{thebibliography}
\bibliographystyle{aasjournal}

\pagebreak

\appendix

\begin{figure}[ht]
\centering
\includegraphics[width=\textwidth]{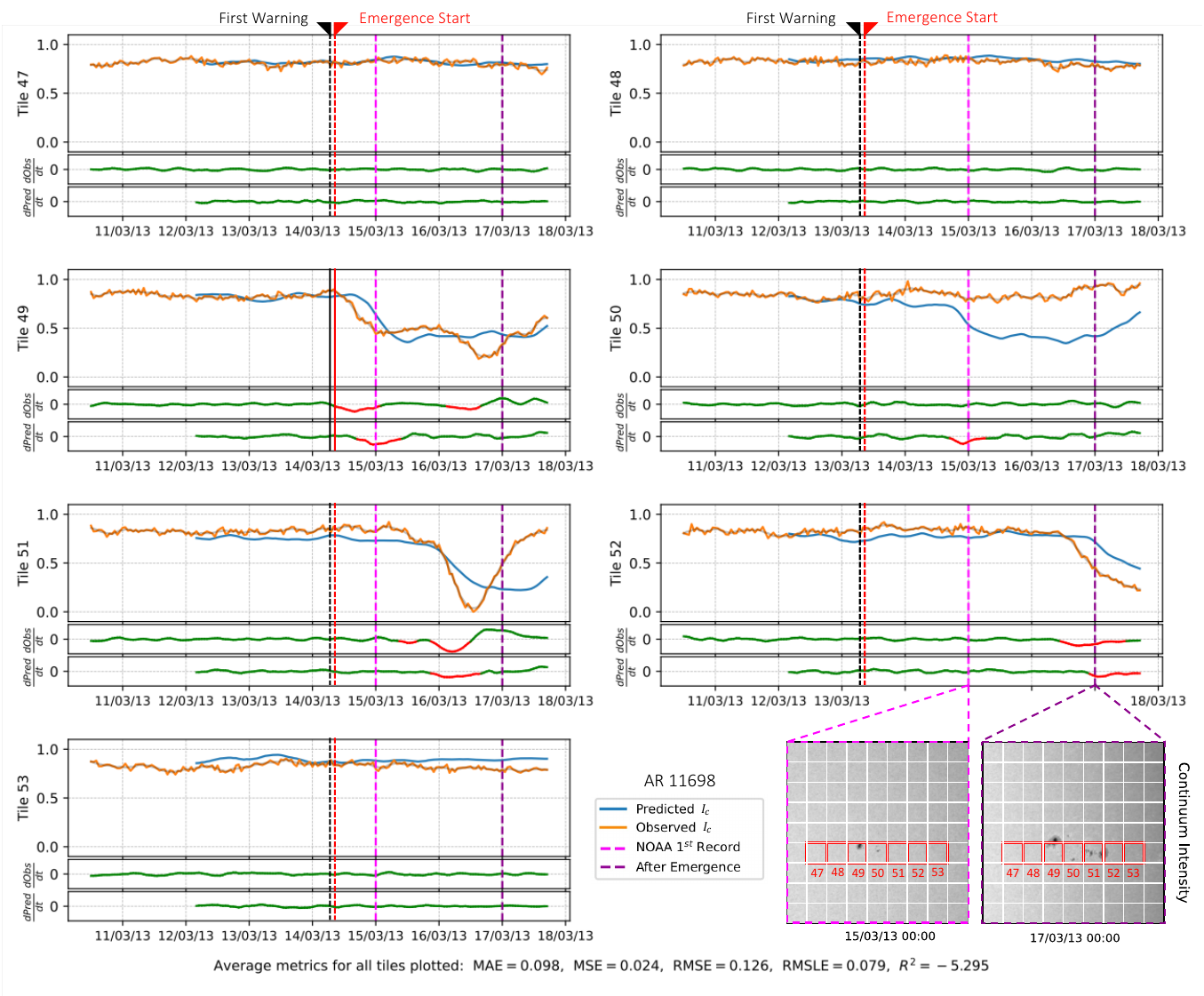}
\caption{Evaluation of Model~8 on predicting variations of the $I_c$ for the selected tiles of AR11698 (tiles 47-53). The tiles' locations are marked in red squares at the bottom right continuum intensity images. Each tile's corresponding observed and predicted mean continuum intensity variations are shown as orange and blue curves. The time-derivatives of the continuum intensity (observed and predicted) are shown color-coded according to the Equation~\ref{eq:criterion} criteria. In red are the emergence periods while in green are the non-emerging (quiet) states. Tiles 49-52 are `active' because a decrease in the continuum intensity is observed, while tile 49 exhibits the first signatures of AR13179 emergence. The rest of the tiles are quiet and correspond to non-emerging states. Vertical dashed lines identify the following moments: NOAA's first record of the active region (magenta), the time two days after NOAA's first record (purple), the time when Model~8 produces its first emergence alarm (First Warning, black) and the time when the observed emergence starts (Emergence Start, red). The tiles for which instead of dashed, the First Warning and Emergence lines are solid, are the tiles in which these events took place.}
\label{fig:results11698}
\end{figure}

\begin{figure}[ht]
\centering
\includegraphics[width=\textwidth]{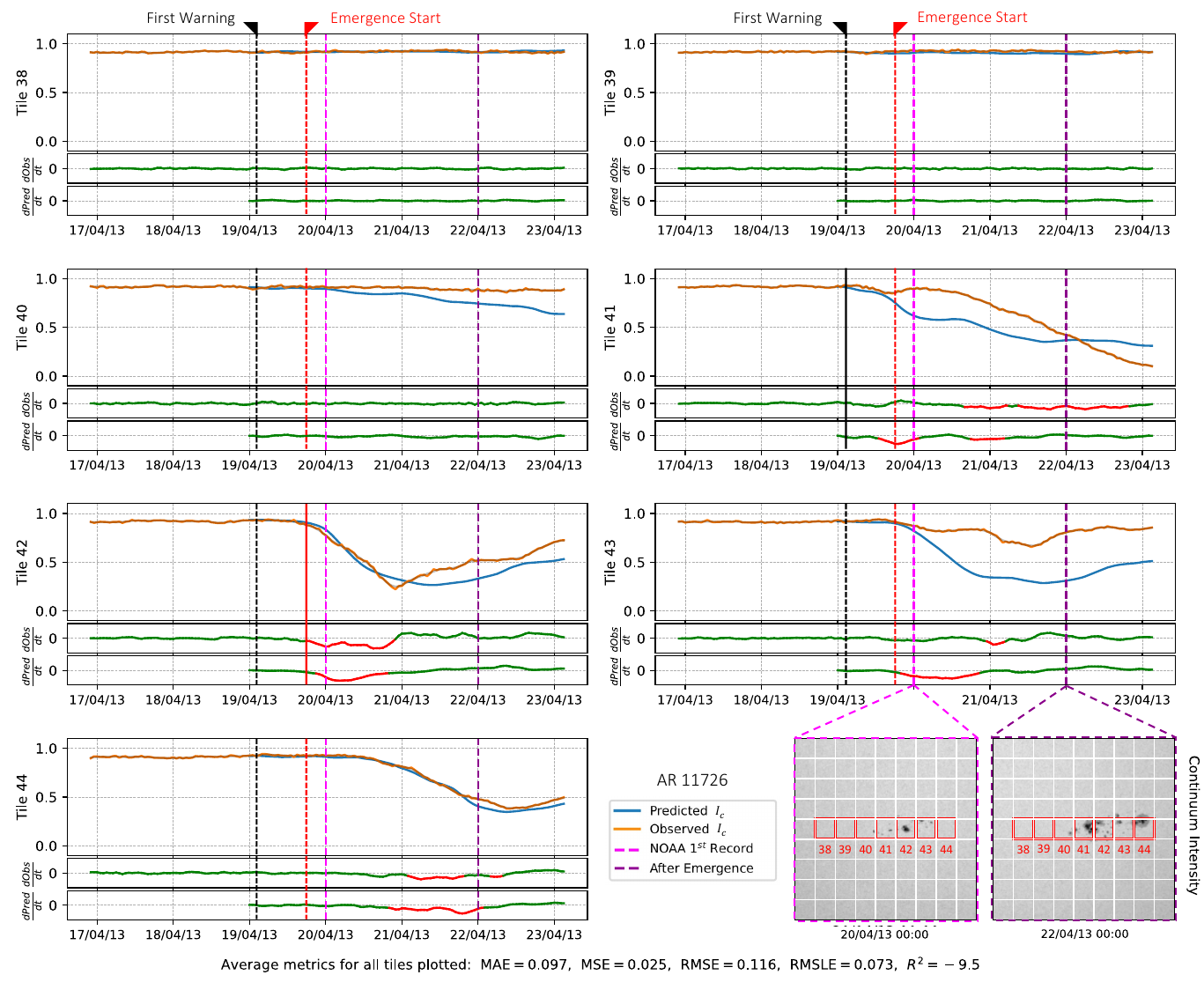}
\caption{Evaluation of Model~8 on predicting variations of the $I_c$ for the selected tiles of AR11726 (tiles 38-44). The tiles' locations are marked in red squares at the bottom right continuum intensity images. Each tile's corresponding observed and predicted mean continuum intensity variations are shown as orange and blue curves. The time-derivatives of the continuum intensity (observed and predicted) are shown color-coded according to the Equation~\ref{eq:criterion} criteria. In red are the emergence periods while in green are the non-emerging (quiet) states. Tiles 41-44 are `active' because a decrease in the continuum intensity is observed, while tile 42 exhibits the first signatures of AR13179 emergence. The rest of the tiles are quiet and correspond to non-emerging states. Vertical dashed lines identify the following moments: NOAA's first record of the active region (magenta), the time two days after NOAA's first record (purple), the time when Model~8 produces its first emergence alarm (First Warning, black) and the time when the observed emergence starts (Emergence Start, red). The tiles for which instead of dashed, the First Warning and Emergence lines are solid, are the tiles in which these events took place.}
\label{fig:results11726}
\end{figure}

\begin{figure}[ht]
\centering
\includegraphics[width=\textwidth]{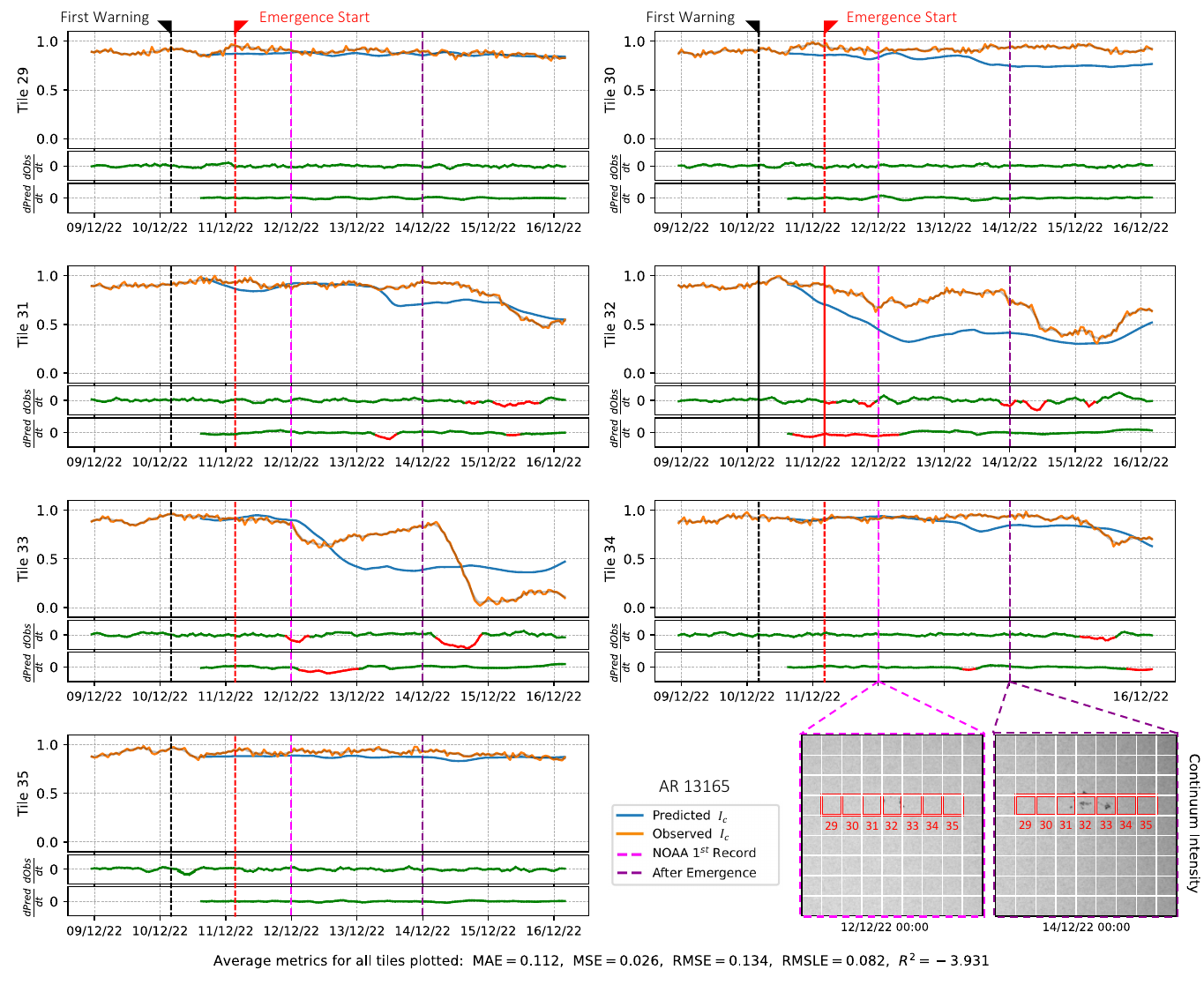}
\caption{Evaluation of Model~8 on predicting variations of the $I_c$ for the selected tiles of AR13165 (tiles 29-35). The tiles' locations are marked in red squares at the bottom right continuum intensity images. Each tile's corresponding observed and predicted mean continuum intensity variations are shown as orange and blue curves. The time-derivatives of the continuum intensity (observed and predicted) are shown color-coded according to the Equation~\ref{eq:criterion} criteria. In red are the emergence periods while in green are the non-emerging (quiet) states. Tiles 31-33 are `active' because a decrease in the continuum intensity is observed, while tile 32 exhibits the first signatures of AR13179 emergence. The rest of the tiles are quiet and correspond to non-emerging states. Vertical dashed lines identify the following moments: NOAA's first record of the active region (magenta), the time two days after NOAA's first record (purple), the time when Model~8 produces its first emergence alarm (First Warning, black) and the time when the observed emergence starts (Emergence Start, red). The tiles for which instead of dashed, the First Warning and Emergence lines are solid, are the tiles in which these events took place.}
\label{fig:results13165}
\end{figure}

\begin{figure}[ht]
\centering
\includegraphics[width=\textwidth]{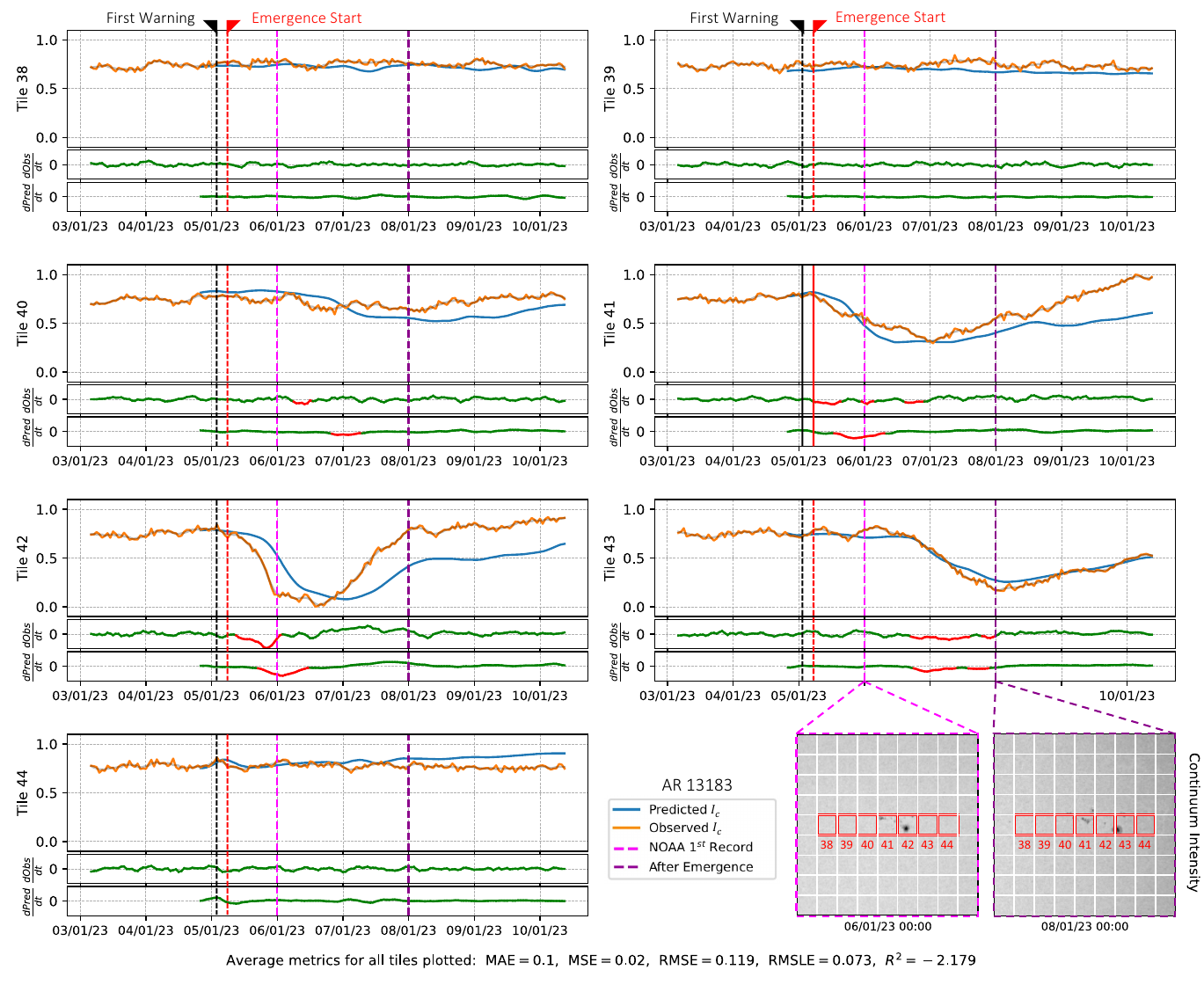}
\caption{Evaluation of Model~8 on predicting variations of the $I_c$ for the selected tiles of AR13183 (tiles 38-44). The tiles' locations are marked in red squares at the bottom right continuum intensity images. Each tile's corresponding observed and predicted mean continuum intensity variations are shown as orange and blue curves. The time-derivatives of the continuum intensity (observed and predicted) are shown color-coded according to the Equation~\ref{eq:criterion} criteria. In red are the emergence periods while in green are the non-emerging (quiet) states. Tiles 40-43 are `active' because a decrease in the continuum intensity is observed, while tile 41 exhibits the first signatures of AR13179 emergence. The rest of the tiles are quiet and correspond to non-emerging states. Vertical dashed lines identify the following moments: NOAA's first record of the active region (magenta), the time two days after NOAA's first record (purple), the time when Model~8 produces its first emergence alarm (First Warning, black) and the time when the observed emergence starts (Emergence Start, red). The tiles for which instead of dashed, the First Warning and Emergence lines are solid, are the tiles in which these events took place.}
\label{fig:results13183}
\end{figure}

\begin{table}[ht]
\label{tab:Training}
\centering
\begin{tabular}{@{}ccccccccccccc@{}}
\toprule
AR\# & First Record & Last Record & $\phi$ & $\lambda_{Carr}$ & $\lambda_{s}$ & $\lambda_{e}$ & $A_s$ & $A_e$ & $A_{max}$ & $A_{max}$ Date & McIntosh & Hale  \\
\midrule
11130 & 2010.11.29 & 2010.12.06 & 12.5 & 330.5 & 0.0 & 94.0 & 60 & 40 & 250 & 2010.12.02 & Dai & $\beta$ \\
11149 & 2011.01.22 & 2011.01.28 & 17.0 & 346.5 & 5.0 & 90.0 & 70 & 0 & 250 & 2011.01.27 & Dso & $\beta$ \\
11158 & 2011.02.12 & 2011.02.21 & -20.0 & 33.5 & -25.0 & 88.0 & 40 & 200 & 620 & 2011.02.17 & Ekc & $\beta \gamma \delta$ \\
11162 & 2011.02.19 & 2011.02.25 & 17.5 & 337.0 & 6.0 & 89.0 & 260 & 0 & 260 & 2011.02.19 & Dai & $\beta \gamma$ \\
11199 & 2011.04.27 & 2011.05.02 & 19.5 & 188.5 & 23.0 & 86.0 & 20 & 210 & 210 & 2011.05.02 & Dso & $\beta$ \\
11327 & 2011.10.21 & 2011.10.28 & -21.0 & 335.5 & -10.0 & 81.0 & 10 & 60 & 200 & 2011.10.25 & Dso & $\beta$ \\
11344 & 2011.11.08 & 2011.11.15 & -19.5 & 102.5 & -6.0 & 86.0 & 0 & 60 & 240 & 2011.11.14 & Esi & $\beta$ \\
11387 & 2011.12.26 & 2011.12.30 & -20.5 & 226.0 & 28.0 & 83.0 & 30 & 170 & 290 & 2011.12.28 & Dki & $\beta \gamma$ \\
11393 & 2012.01.06 & 2012.01.12 & 17.0 & 55.5 & 3.0 & 84.0 & 40 & 560 & 560 & 2012.01.12 & Eko & $\beta$ \\
11416 & 2012.02.10 & 2012.02.18 & -16.5 & 288.5 & -24.0 & 84.0 & 90 & 110 & 400 & 2012.02.12 & Dhi & $\beta \gamma$ \\
11422 & 2012.02.20 & 2012.02.27 & 15.0 & 177.0 & -2.0 & 90.0 & 60 & 150 & 330 & 2012.02.22 & Dsi & $\beta$ \\
11455 & 2012.04.12 & 2012.04.18 & 6.0 & 208.0 & -8.0 & 78.0 & 30 & 30 & 200 & 2012.04.17 & Fsi & $\beta$ \\
11619 & 2012.11.18 & 2012.11.25 & 11.5 & 171.5 & -16.0 & 78.0 & 90 & 100 & 200 & 2012.11.20 & Dsi & $\beta$ \\
11640 & 2013.01.01 & 2013.01.08 & 28.0 & 321.0 & -20.0 & 87.0 & 40 & 300 & 360 & 2013.01.07 & Eho & $\beta$ \\
11660 & 2013.01.20 & 2013.01.27 & 12.0 & 64.5 & -12.0 & 81.0 & 50 & 100 & 220 & 2013.01.26 & Dao & $\beta$ \\
11678 & 2013.02.19 & 2013.02.24 & 9.0 & 68.0 & 26.0 & 93.0 & 30 & 350 & 470 & 2013.02.22 & Dkc & $\beta \gamma \delta$ \\
11682 & 2013.02.26 & 2013.03.06 & -18.0 & 298.5 & -9.0 & 94.0 & 30 & 60 & 240 & 2013.03.02 & Dai & $\beta \gamma$ \\
11765 & 2013.06.04 & 2013.06.13 & 9.0 & 52.5 & -17.0 & 78.0 & 30 & 30 & 210 & 2013.06.10 & Dai & $\beta$ \\
11768 & 2013.06.13 & 2013.06.18 & -11.5 & 354.5 & 19.0 & 85.0 & 50 & 260 & 320 & 2013.06.15 & Dko & $\beta$ \\
11776 & 2013.06.19 & 2013.06.26 & 11.0 & 252.0 & -5.0 & 88.0 & 10 & 30 & 220 & 2013.06.23 & Dai & $\beta \gamma$ \\
11916 & 2013.12.05 & 2013.12.13 & -13.5 & 167.0 & -19.0 & 90.0 & 10 & 120 & 350 & 2013.12.11 & Ekc & $\beta \gamma$ \\
11928 & 2013.12.18 & 2013.12.25 & -15.5 & 6.0 & -8.0 & 87.0 & 130 & 130 & 460 & 2013.12.22 & Ekc & $\beta \gamma$ \\
12036 & 2014.04.14 & 2014.04.23 & -17.5 & 246.0 & -25.0 & 92.0 & 20 & 90 & 510 & 2014.04.18 & Dhc & $\beta \gamma$ \\
12051 & 2014.05.02 & 2014.05.07 & -9.0 & 58.0 & 22.0 & 92.0 & 10 & 170 & 310 & 2014.05.04 & Dkc & $\beta \gamma \delta$ \\
12085 & 2014.06.07 & 2014.06.16 & -19.5 & 255.0 & -24.0 & 97.0 & 20 & 0 & 840 & 2014.06.10 & Ekc & $\beta \gamma \delta$ \\
12089 & 2014.06.12 & 2014.06.20 & 18.0 & 197.0 & -16.0 & 92.0 & 20 & 120 & 270 & 2014.06.17 & Dkc & $\beta \gamma \delta$ \\
12144 & 2014.08.15 & 2014.08.20 & -17.0 & 104.0 & 19.0 & 87.0 & 20 & 100 & 220 & 2014.08.18 & Dsi & $\beta \gamma$ \\
12175 & 2014.09.26 & 2014.10.02 & 16.5 & 262.0 & 12.0 & 90.0 & 80 & 140 & 500 & 2014.09.30 & Ekc & $\beta \gamma \delta$ \\
12203 & 2014.11.02 & 2014.11.09 & 12.5 & 114.0 & -8.0 & 84.0 & 30 & 20 & 200 & 2014.11.04 & Dao & $\beta$ \\
12257 & 2015.01.07 & 2015.01.15 & 6.0 & 320.0 & -12.0 & 93.0 & 40 & 120 & 480 & 2015.01.10 & Dki & $\beta \delta$ \\
12331 & 2015.04.22 & 2015.04.29 & -9.5 & 21.0 & -10.0 & 87.0 & 30 & 210 & 240 & 2015.04.27 & Dai & $\beta \gamma$ \\
12494 & 2016.02.05 & 2016.02.13 & -13.0 & 164.0 & -9.0 & 96.0 & 150 & 10 & 270 & 2016.02.07 & Dki & $\beta \delta$ \\
12659 & 2017.05.22 & 2017.05.30 & 13.5 & 41.0 & -26.0 & 86.0 & 40 & 180 & 220 & 2017.05.29 & Dao & $\beta$ \\
12778 & 2020.10.27 & 2020.11.01 & -20.5 & 85.5 & 17.0 & 80.0 & 80 & 100 & 300 & 2020.10.30 & Eko & $\beta$ \\
12864 & 2021.09.04 & 2021.09.12 & 0.5 & 253.0 & -22.0 & 90.0 & 30 & 50 & 220 & 2021.09.08 & Cso & $\beta$ \\
12877 & 2021.09.27 & 2021.10.04 & -19.0 & 328.0 & 0.0 & 93.0 & 10 & 30 & 200 & 2021.10.01 & Dao & $\beta \gamma$ \\
12900 & 2021.11.27 & 2021.12.03 & -27.5 & 255.0 & 12.0 & 90.0 & 50 & 0 & 240 & 2021.11.30 & Dao & $\beta$ \\
12929 & 2022.01.14 & 2022.01.21 & 8.0 & 323.5 & -9.0 & 88.0 & 10 & 80 & 310 & 2022.01.17 & Dki & $\beta$ \\
13004 & 2022.05.03 & 2022.05.10 & -14.5 & 324.5 & -12.0 & 85.0 & 30 & 180 & 500 & 2022.05.06 & Dkc & $\beta \delta$ \\
13085 & 2022.08.22 & 2022.08.30 & 30.0 & 290.0 & -16.0 & 90.0 & 50 & 60 & 280 & 2022.08.26 & Dko & $\beta$ \\
13098 & 2022.09.09 & 2022.09.17 & 16.5 & 51.0 & -17.0 & 91.0 & 10 & 540 & 860 & 2022.09.15 & Ehc & $\beta \gamma$ \\
\bottomrule
\end{tabular}
\caption{Summary of the assigned NOAA active region number (AR\#), the date of NOAA's first (First Record) and last (Last Record) record, the constant latitude in which the AR emerged ($\phi$), the Carrington longitude ($\lambda_{Carr}$), the starting ($\lambda_{s}$) and ending ($\lambda_{e}$) longitude, the starting ($A_s$), ending ($A_e$) and maximum ($A_{max}$) area of the AR in millionths of the hemisphere, the date of the maximum area ($A_{max}$ Date), McIntosh classification, and Hale classification for the ARs used for training in this work.}
\end{table}

\end{document}